\documentclass[11pt,oneside,letterpaper]{article}
\pdfoutput=1
\usepackage{amssymb}
\usepackage{amsmath}
\usepackage[dvips]{graphicx}
\usepackage{setspace}
\usepackage{amsfonts}
\usepackage{fancyhdr}
\usepackage{xcolor}
\usepackage{graphicx}
\usepackage{rotating}
\usepackage{comment}
\usepackage{color}
\usepackage{cite}
\usepackage{braket}
\definecolor{darkgreen}{rgb}{0,0.5,0}
\definecolor{darkblue}{rgb}{0,0,0.6}
\definecolor{purple}{rgb}{0.4,.2,0.7}
\newcommand{\p}{\partial}

\newcommand{\be}{\begin{equation}}
\newcommand{\ee}{\end{equation}}

\usepackage[colorlinks=true,citecolor=darkgreen,linkcolor=black,urlcolor=purple]{hyperref}

\usepackage{pdfsync}

\makeatletter
\newcommand*{\defeq}{\mathrel{\rlap{%
                     \raisebox{0.3ex}{$\m@th\cdot$}}%
                     \raisebox{-0.3ex}{$\m@th\cdot$}}%
                     =} 
\makeatother

\DeclareMathOperator{\Tr}{Tr}
\def\be{\begin{eqnarray}}
\def\ee{\end{eqnarray}}

\newcommand{\la}{\langle}

\newcommand{\tr}{\textrm{Tr}\,}

\newcommand{\bea}{\begin{eqnarray}}
\newcommand{\eea}{\end{eqnarray}}
\def\ben{\begin{equation}}
\def\een{\end{equation}}
\def\half{{\textstyle{\frac{1}{2}}}}

    \let\p=\phi \let\r=v

\let\la=\label

\def\be{\begin{equation}}
\def\ee{\end{equation}}
\def\ba{\begin{array}}
\def\ea{\end{array}}

\def\ba#1\ea{\begin{align}#1\end{align}}
\def\bs#1\es{\begin{split}#1\end{split}}

\renewcommand{\p}{\partial}

\allowdisplaybreaks  

\numberwithin{equation}{section}

\def\nref#1{(\ref{#1})}
\def \la {\label}   
\def \be {\begin{equation}}
\def \ee {\end{equation}}
\def \half {{1\over 2}}	
\def \JM#1 {{\color{blue}  JM: #1 }}
\def \AAl#1 {{\color{red}  AA: #1 }}

\begin{document}
\onehalfspacing

\begin{center}

~
\vskip5mm

{\LARGE  {
Replica Wormholes and the Entropy of Hawking Radiation \\
\ \\
}}

\vskip10mm

Ahmed Almheiri,$^1$\ \ Thomas Hartman,$^{2}$\ \ Juan Maldacena,$^{1}$\\  Edgar Shaghoulian,$^{2}$\ \  and Amirhossein Tajdini$^{2}$

\vskip5mm

{\it $^1$ Institute for Advanced Study, Princeton, New Jersey, USA } \\
\vskip5mm
{\it $^2$ Department of Physics, Cornell University, Ithaca, New York, USA
} 

\vskip5mm

\end{center}

\vspace{4mm}

\begin{abstract}
\noindent

The information paradox can be realized in anti-de Sitter spacetime joined to a Minkowski region. In this setting, we show that the large discrepancy between the von Neumann entropy as calculated by Hawking and the requirements of unitarity is fixed by including new saddles in the gravitational path integral. These saddles arise in the replica method as complexified wormholes connecting different copies of the black hole. As the replica number $n \to 1$, the presence of these wormholes leads to the island rule for the computation of the fine-grained gravitational entropy. 
We discuss these replica wormholes explicitly in two-dimensional Jackiw-Teitelboim gravity coupled to matter.

 \end{abstract}

\pagebreak
\pagestyle{plain}

\setcounter{tocdepth}{2}
{}
\vfill
\tableofcontents

\newpage

\section{Introduction}


Hawking famously noted that the process of black hole formation and evaporation seems to create entropy \cite{Hawking:1976ra}. We can form a black hole from a pure state. The formation of the black hole horizon leaves an inaccessible region behind, and the entanglement of quantum fields across the horizon is responsible for the thermal nature of the Hawking radiation as well as its growing entropy. 

A useful diagnostic for information loss is the fine-grained (von Neumann) entropy of the Hawking radiation, $S_R = -\tr \rho_R \log \rho_R$, where $\rho_R$ is the density matrix of the radiation. This entropy initially increases, because the Hawking radiation is entangled with its partners in the black hole interior. But if the evaporation is unitary, then it must eventually fall back to zero following the Page curve \cite{Page:1993wv,Page:2013dx}.  On the other hand, Hawking's calculation predicts an entropy that rises monotonically as the black hole evaporates.

\begin{figure}
\begin{center}
\includegraphics[scale=0.8]{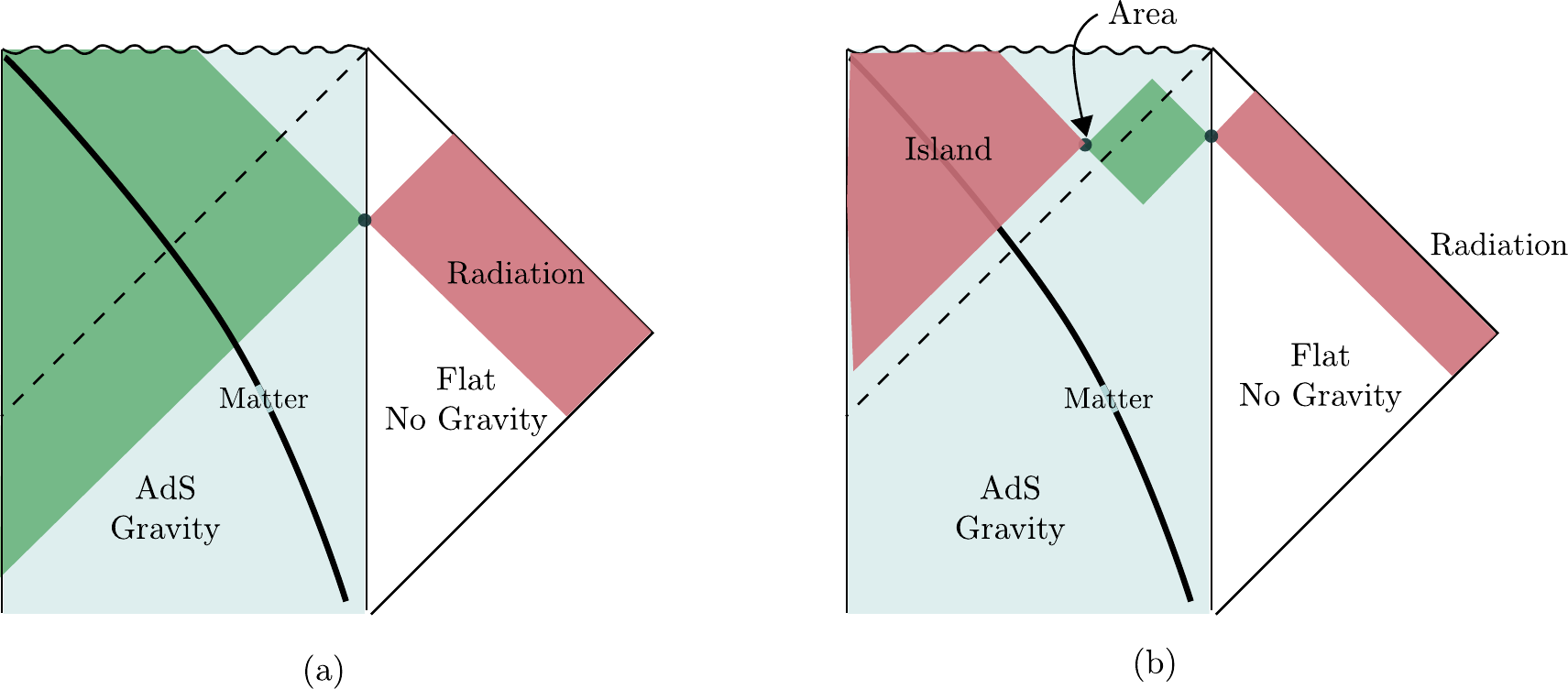}
\end{center}
\caption{  We display an evaporating black hole. The vertical line separates a region on the left where gravity is dynamical from a region on the right where we can approximate it as not being dynamical. The black hole is evaporating into this second region. In red we see the regions associated to the computation of the entropy of radiation and in green the regions computing the entropy of the black hole. (a) Early times. (b) Late times, where we have an island. } 
\label{fig:QESreview}
\end{figure}

Hawking's computation of the entropy seems straightforward. It can be done far from the black hole where the
effects of quantum gravity are small, so it is unclear what could have gone wrong. 
An answer to this puzzle was recently proposed \cite{Penington:2019npb,Almheiri:2019psf,Almheiri:2019hni}  (see also 
\cite{Mertens:2019bvy,Akers:2019wxj,Moitra:2019xoj,Almheiri:2019psy,Fu:2019oyc,Zhang:2019fcy,Akers:2019nfi,Almheiri:2019yqk,Rozali:2019day,Chen:2019uhq,Bousso:2019ykv,Jafferis:2019wkd,Blommaert:2019wfy}). The proposal is that Hawking used the wrong formula for computing the entropy. As the theory is coupled to gravity, we should use the proper gravitational formula for entropy: the gravitational fine-grained entropy formula studied by Ryu and Takayanagi \cite{Ryu:2006bv} and extended in \cite{Hubeny:2007xt,Faulkner:2013ana,Engelhardt:2014gca}, also allowing for 
spatially disconnected regions, called ``islands,'' see 
 figure \ref{fig:QESreview}.  Even though the radiation lives in a region where the gravitational effects are small, the fact that we are describing a state in a theory of gravity 
 implies that we should use the gravitational formula for the entropy, including the island rule.

In this paper we consider a version of the information paradox formulated recently in \cite{Penington:2019npb,Almheiri:2019psf} (see also \cite{Mathur:2014dia})
where a black hole in anti-de Sitter spacetime radiates into an attached Minkowski region.  We show that the first principles computation of the fine-grained entropy using the gravitational path integral description   receives large corrections from non-perturbative effects. The effects come from new saddles in the gravitational path integral --- replica wormholes --- that dominate over the standard Euclidean black hole saddle, and lead to a fine-grained entropy consistent with unitarity.

We will discuss the saddles explicitly only in some simple examples related to the information paradox for eternal black holes in two-dimensional Jackiw-Teitelboim (JT) gravity \cite{Jackiw:1984je,Teitelboim:1983ux,Almheiri:2014cka}, reviewed below, but we can nonetheless compute the effect on the fine-grained entropy more generally.
The same answer for the entropy was obtained holographically in \cite{Almheiri:2019hni,Chen:2019uhq,Rozali:2019day}. 
Our goal is to provide a direct, bulk derivation   without using holography.

To summarize our approach briefly, we will revisit the calculation of the von Neumann entropy of radiation outside a black hole in AdS glued to flat space, using the replica method. We introduce $n$ copies of the original black hole, analytically continue to non-integer $n$, and compute the von Neumann entropy as $S_R = - \p_n \tr (\rho_R)^n|_{n = 1}$. Since the theory is coupled to gravity, we must do the gravitational path integral  to calculate $\tr (\rho_R)^n$. Under our assumptions about the matter content, this path integral is dominated by a saddlepoint. There is one obvious saddle, in which the geometry is $n$ copies of the original black hole; this saddle leads to the standard Hawking result for the von Neumann entropy, \textit{i.e.}, the entropy of quantum fields in a fixed curved spacetime, see figure \ref{fig:2wormhole}(a).

 There is, however, another class of saddles in which the different replicas are connected by a new geometry. These are the replica wormholes, see figure \ref{fig:2wormhole}(b), \ref{fig:manyreplicas}. In the examples we consider, whenever the Hawking-like calculation leads to an entropy in tension with unitarity, the replica wormholes start to dominate the gravitational path integral, and resolve the tension.

Our use of the replica trick in a theory coupled to gravity closely parallels the derivation of   the Ryu-Takayanagi formula  and its generalizations \cite{Lewkowycz:2013nqa,Faulkner:2013ana,Dong:2016hjy,Dong:2017xht}.

In the rest of the introduction we summarize the main idea in more detail.

Similar ideas are explored independently in a paper by Penington, Shenker, Stanford, and Yang \cite{Penington:2019kki}.

\subsection{The island rule for computing gravitational von Neumann entropies}

We begin by reviewing the recent progress on the information paradox in AdS/CFT  \cite{Penington:2019npb,Almheiri:2019psf}.

The classic information paradox is difficult to study in AdS/CFT, because large black holes do not evaporate. Radiation bounces off the AdS boundary and falls back into the black hole.  For this reason, until recently, most discussions of the information paradox in AdS/CFT have focused on exponentially small effects, such as the late-time behavior of boundary correlation functions \cite{Maldacena:2001kr,Saad:2018bqo,Saad:2019lba,Saad:2019pqd}.

In contrast, the discrepancy in the Page curve is a large, $O(1/G_N)$, effect. This classic version of the information paradox can be embedded into AdS/CFT by coupling AdS to an auxiliary system that absorbs the radiation, allowing the black hole to evaporate \cite{Penington:2019npb,Almheiri:2019psf} (see also \cite{Rocha:2008fe, Mertens:2019bvy, Engelsoy:2016xyb}). This is illustrated in fig.~\ref{fig:QESreview} in the case where the auxiliary system is half of Minkowski space, glued to the boundary of AdS. There is no gravity in the Minkowski region, where effectively $G_N \to 0$, but radiation into matter fields is allowed to pass through the interface.

In this setup, the Page curve of the black hole was calculated   in 
\cite{Penington:2019npb,Almheiri:2019psf}. It is important to note that this calculation gives the Page curve of the black hole, not the radiation, which is where the paradox lies; we return to this momentarily.
The entropy of the black hole is given by the generalized entropy of the quantum extremal surface (QES) \cite{Engelhardt:2014gca}, which is a quantum-corrected  Ryu-Takayanagi (or Hubeny-Rangamani-Takayanagi) surface \cite{Ryu:2006bv,Hubeny:2007xt}. According to the QES proposal, the von Neumann entropy of the black hole is
\be\label{introsbh}
S_B = \mbox{ext}_Q \left[ \frac{\mbox{Area}(Q)}{4G_N} + S_{\rm matter}(B) \right]
\ee
where $Q$ is the quantum extremal surface, and $B$ is the region between $Q$ and the AdS boundary. $S_{\rm matter}$ denotes the von Neumann entropy of the  quantum field theory (including perturbative gravitons) calculated in the fixed background geometry. The extremization is over the choice of surface $Q$. If there is more than one extremum, then $Q$ is the surface with minimal entropy. For dilaton gravity in AdS$_2$, $Q$ is a point, and its `area' means the value of the dilaton.

The black hole Page curve  is the function $S_B(t)$, where $t$ is the time on the AdS boundary where $B$ is anchored. 
It depends on time  because the radiation can cross into the auxiliary system. 
It behaves as expected: it grows at early times, then eventually falls back to zero \cite{Penington:2019npb,Almheiri:2019psf}. A crucial element of this analysis is that at late times, the dominant quantum extremal surface sits near the black hole horizon, as in fig.~\ref{fig:QESreview}.

This does not resolve the Hawking paradox, 
 which involves the radiation entropy $S_{\rm matter}(R)$, where $R$ is a region outside the black hole containing the radiation that has come out. 
Clearly the problem is that neither $R$ nor $B$ includes the region $I$ behind the horizon, called the island, 
see figure \ref{fig:QESreview}. The state of the quantum fields on $R \cup B$ is apparently not pure, and, apparently  $S_R \neq S_B$. Only if we assume unitarity, or related holographic input such as entanglement wedge reconstruction \cite{Penington:2019npb}, can we claim that the QES computes the entropy of the radiation.  
It does, however, tell us what to aim for in a unitary theory.

 With this motivation, in \cite{Almheiri:2019hni}, the evaporating black hole in Jackiw-Teitelboim (JT) gravity in AdS$_2$ was embedded into a holographic theory in one higher dimension. The AdS$_2$ black hole lives on a brane at the boundary of AdS$_3$, similar to a Randall-Sundrum model \cite{Randall:1999vf,Karch:2000ct}, with JT gravity on the brane (see also \cite{Almheiri:2019psy} for an analogous construction on an AdS$_4$ boundary of AdS$_5$). In this setup, \cite{Almheiri:2019hni} derived the QES prescription \textit{for the radiation} using AdS$_3$ holography. It was found that the von Neumann entropy of the radiation in region $R$, computed holographically in AdS$_3$, agrees with the black hole entropy in \eqref{introsbh}. This led to the conjecture that in a system coupled to gravity, the ordinary calculation of von Neumann entropy should be supplemented by the contribution from ``islands" according to the following rule:
\be \la{IslRule}
S(\rho_R) = \mbox{ext}_Q \left[ \frac{\mbox{Area}(Q)}{4G_N} + S(\tilde{\rho}_{I \cup R}) \right] \ ,
\ee
up to subleading corrections. Here $\rho_R$ is the density matrix of the region $R$ in the full theory coupled to quantum gravity, and $\tilde{\rho}_{I \cup R}$ is the density matrix of the state prepared via the semi-classical path integral on the Euclidean black hole saddle.
This is equal to \eqref{introsbh}, since the quantum fields are pure on the full Cauchy slice $I \cup B \cup R$. Thus the tension with unitarity is resolved within three-dimensional holography.

In this paper we explain how the surprising island rule \nref{IslRule} follows from the standard rules for computing gravitational fine-grained entropy,   without appealing to   higher dimensional holography.

\subsection{Two dimensional eternal black holes and the information paradox}

We consider an AdS$_2$   JT gravity theory coupled to a 2d CFT. This CFT also lives in   non-gravitational Minkowski regions, and has transparent boundary conditions at the $AdS$ boundary. 
 The dilaton goes to infinity at the AdS$_2$ boundary so it is consistent to freeze gravity on the outside \cite{Engelsoy:2016xyb,Almheiri:2019psf}. We will assume that the matter CFT has a large central charge $c \gg 1$, but we will not assume that it is holographic, as all our calculations are done directly in the 2d theory. For example it could be $c$ free bosons. Taking the central charge large is to suppress the quantum fluctuations of the (boundary) graviton relative to the matter sector. 

This simple model of an $AdS_2$ black hole glued to flat space can be directly applied to certain four dimensional 
 black holes. For example, for the near extremal magnetically charged black holes discussed in \cite{Maldacena:2018gjk}, at low temperatures we can approximate the dynamics as an $AdS_2$ region joined to a flat space region, and the light fields come from effectively two dimensional fields moving in the radial and time direction that connect the two regions. 

\begin{figure}
\begin{center}
\includegraphics[scale=0.6]{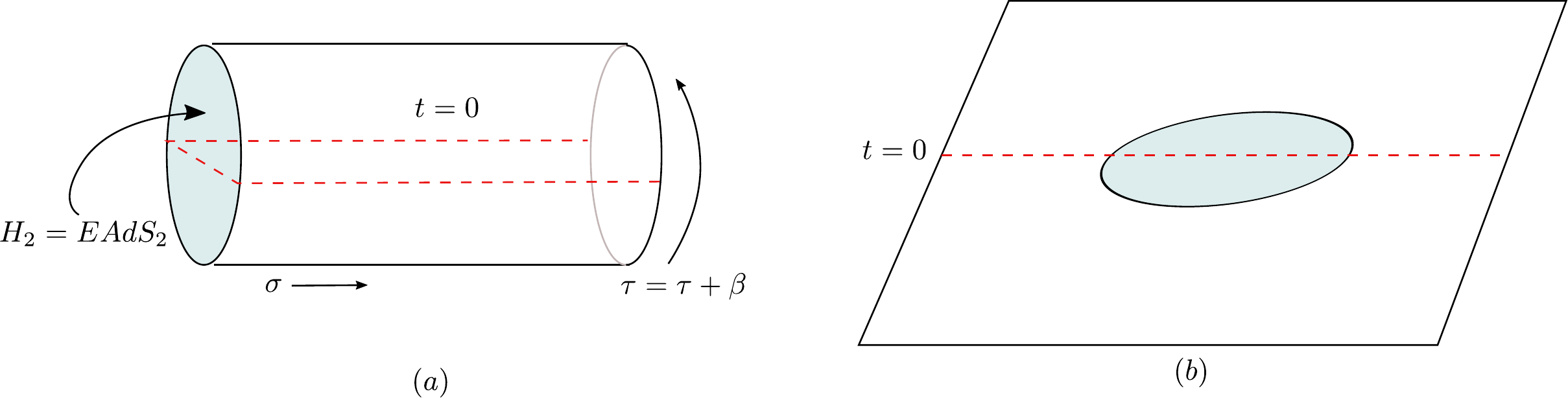}
\end{center}
\caption{We prepare the combined thermofield double state of the black hole and radiation using a Euclidean path integral. These are two pictures for the combined geometry. In   (b) we have represented the outside cylinder as the outside of the disk. By cutting along the red dotted line, we get our desired thermofield double initial state that we can then use for subsequent Lorentzian evolution (forwards or backwards in time) to get 
the diagram in figure \ref{fig:eternalBH-lorentzian}.  } 
\label{fig:TFDpreparation}
\end{figure}

\begin{figure}
\begin{center}
\includegraphics[scale=0.6]{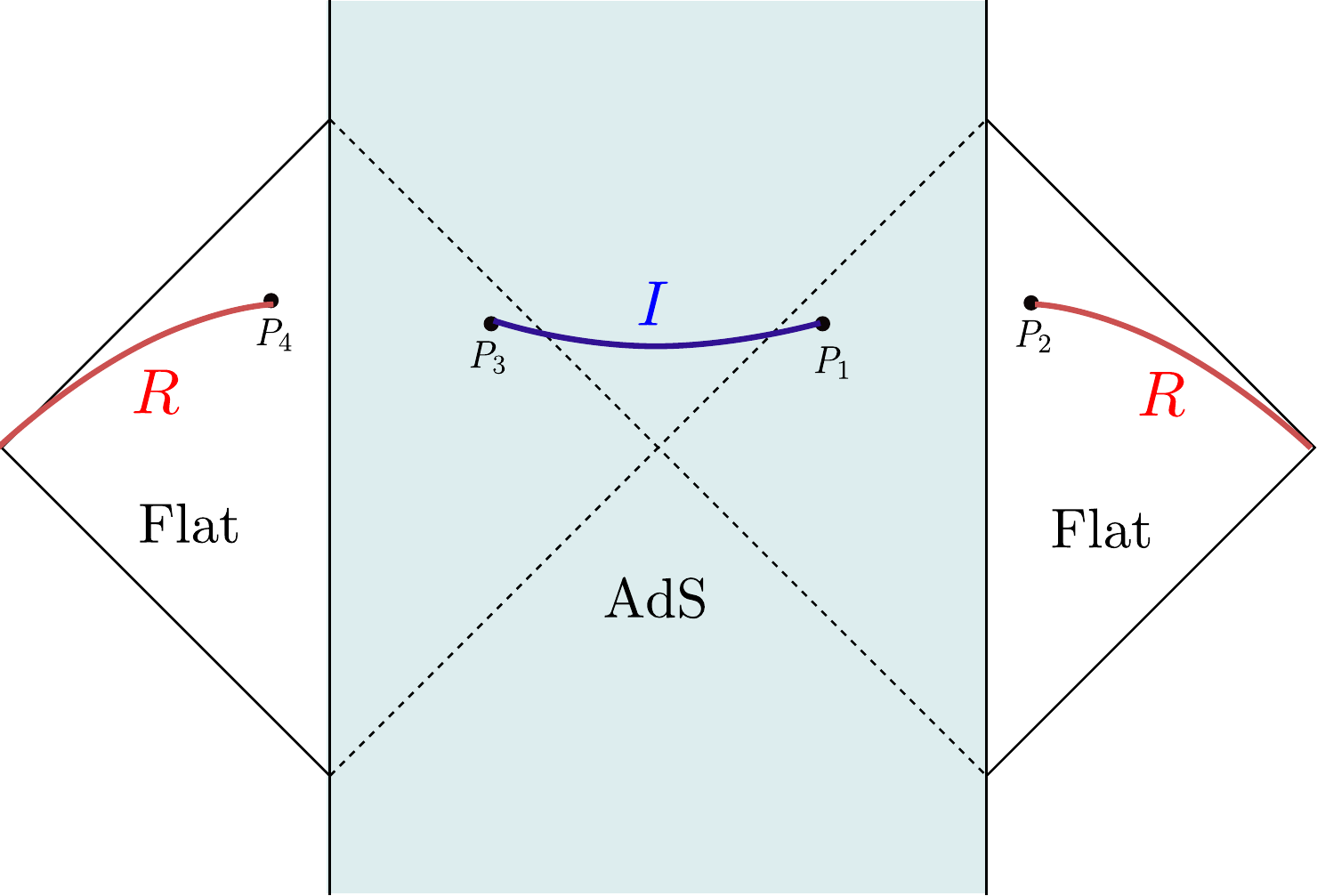}
\end{center}
\caption{Eternal black hole in AdS$_2$, glued to Minkowski space on both sides. Hawking radiation is collected in region $R$, which has two disjoint components. Region $I$ is the island. The shaded region is coupled to JT gravity. 
} 
\label{fig:eternalBH-lorentzian}
\end{figure}

We will consider a simple initial state which is the thermofield double state for the black hole plus radiation. 
This state is prepared by a simple Euclidean path integral, see figure \ref{fig:TFDpreparation}. The resulting
Lorentzian geometry is shown in figure \ref{fig:eternalBH-lorentzian}.

\begin{figure}
\begin{center}
\includegraphics[scale=0.7]{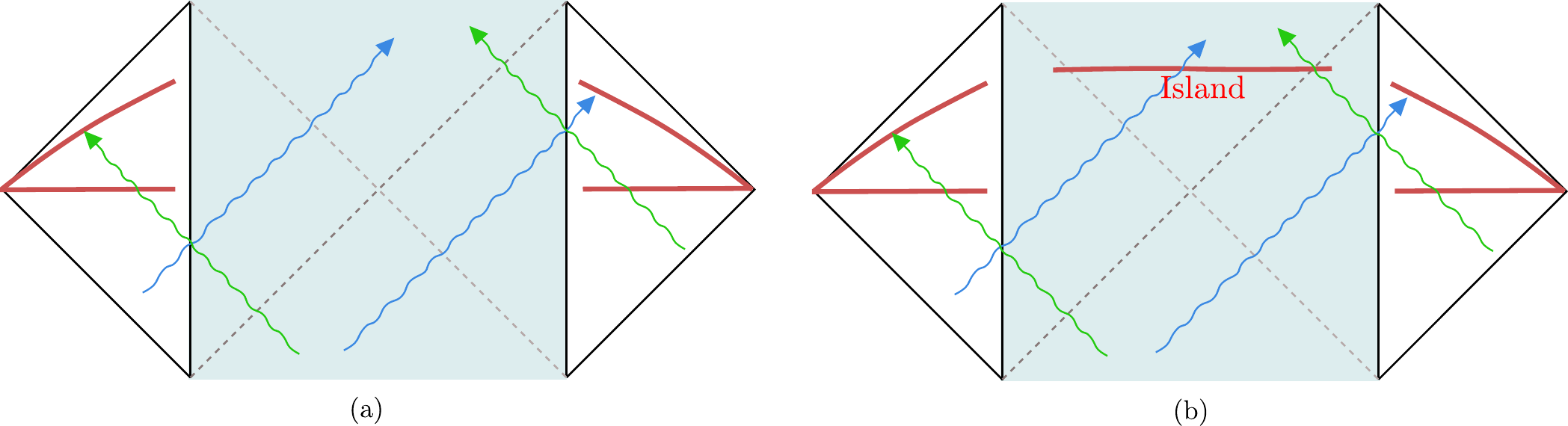}
\end{center}
\caption{(a) Growing entropy for the radiation for an eternal black hole plus radiation in the thermofield double state. We draw two instants in time. The particles with the same color are entangled. They do not contribute to the entanglement of the radiation region (indicated in red) at $t=0$ but they do contribute at  a later value of $t$. (b) When the island is included the entanglement ceases to grow, because now   both entangled modes mentioned above are included in $I \cup R$.    } 
\label{fig:Cartoon}
\end{figure}

\begin{figure}
\begin{center}
\includegraphics[scale=1]{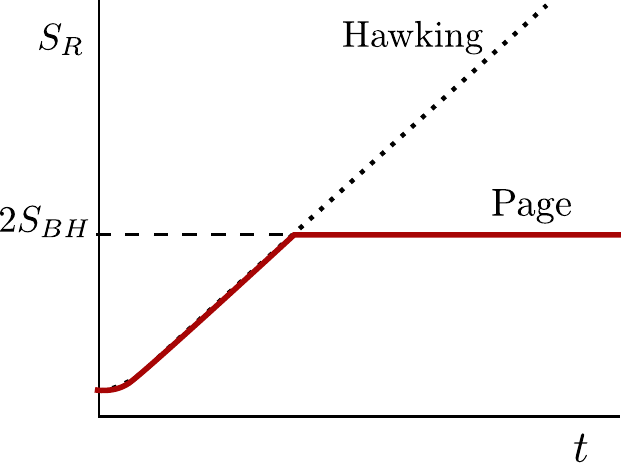}
\end{center}
\caption{  Page curve for the entropy of the radiation, for the model in fig. \ref{fig:eternalBH-lorentzian}. The dotted line is the growing result given by the Hawking computation, and the entropy calculated from the other saddle is dashed. The minimum of the two is the Page curve for this model.     } 
\label{Page}
\end{figure}

Despite its simplicity, this setup exhibits Hawking's information paradox, and the corresponding puzzle with the Page curve \cite{Page:1993wv,Page:2013dx}. To reach a paradox, we collect Hawking radiation in region $R$ in figure
\ref{fig:eternalBH-lorentzian}. As a function of time, $R$ moves upward on both sides of the Penrose diagram, so this is not a symmetry. Indeed, the von Neumann entropy of the radiation as calculated by Hawking, $S_{\rm matter}(R(t))$, grows linearly with time, see fig.~\ref{Page}. 
 The origin of this growth is the following. At $t=0$ the radiation modes on the left are entangled with modes on the right. However, as time progresses some of these modes fall into the black holes, others are replaced by black hole modes, see figure 
\ref{fig:Cartoon}. 

If this growth were to continue forever,  it would become larger  than the 
 Bekenstein-Hawking entropies of the two black holes, and this is a contradiction.  See a related discussion 
 of the critically illuminated black hole in flat spacetime in \cite{Fiola:1994ir}. 
 
In a unitary theory, $S_R(t)$ should saturate at around the twice the  Bekenstein-Hawking entropy of each black hole, see figure \ref{Page}. This was confirmed using the island rule in \cite{Almheiri:2019yqk}.

\subsection{Replica wormholes to the rescue}

To reproduce the unitary answer directly from a gravity calculation, we will use the replica method to compute the von Neumann entropy of region $R$. The saddles relevant to the unitary Page curve will ultimately be complex solutions of the gravitational equations. The idea is to do Euclidean computations and then analytically continue to Lorentzian signature.

\begin{figure} 
\begin{center}

\includegraphics[scale=.9]{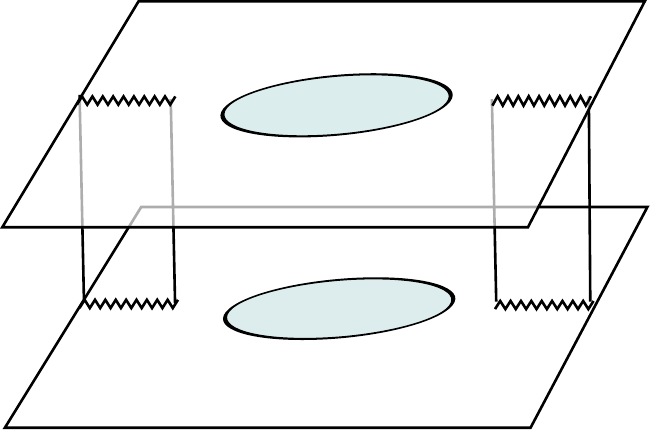} \hspace{20mm}
\includegraphics[scale=.9]{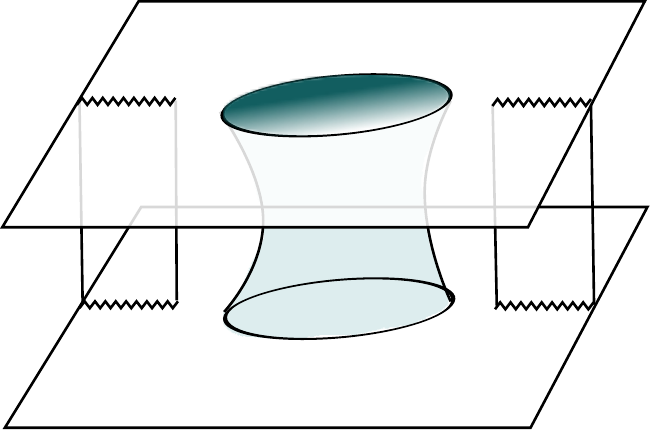}  

 (a) ~~~~~~~~~~~~~~~~~~~~~~~~~~~~~~~~~~~~~~~~~~~~~~~~~~~~~~~~~~~~~~(b)
	\end{center}
\caption{\small Two different saddlepoint contributions to the two-replica path integral in the presence of gravity in the shaded region. On the left the replicas are sewn together along the branch points, outside of the shaded region, as we would do in an ordinary quantum field theory calculation. These will give the standard QFT answer, as computed by Hawking, which can lead to a paradox. On the right we have a saddle where gravity dynamically glues together the shaded regions. This is the replica wormhole. In the examples considered in this paper, this saddle dominates in the relevant kinematics, leading to a Page curve consistent with unitarity. \label{fig:2wormhole}}
\end{figure}

Consider $n=2$ replicas. The replica partition function $\tr (\rho_R)^2$ is computed by a Euclidean path integral on two copies of the Euclidean system, with the matter sector sewed together along the cuts on region $R$. Since we are doing a gravitational path integral, we do not specify the geometry in the gravity region; we only fix the boundary conditions at the edge. Gravity then fills in the geometry  dynamically, see fig.~\ref{fig:2wormhole}. 

\begin{figure}
\begin{center}
\includegraphics[scale=0.4]{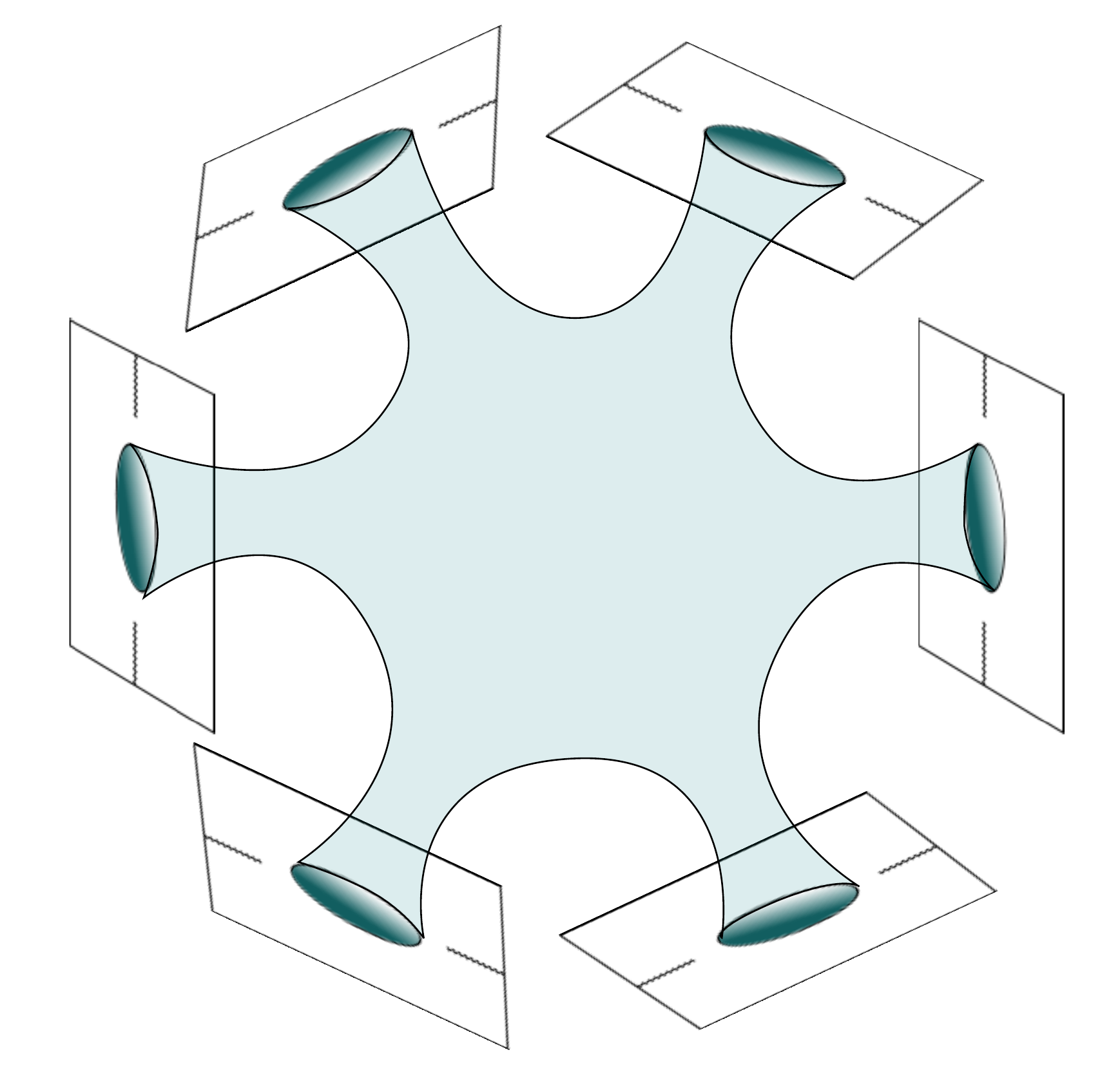}
\end{center}
\caption{\small Topology of a replica wormhole with $n=6$. The sheets are also glued together cyclically along the cuts in the matter region. \label{fig:manyreplicas}}
\end{figure}

We consider two different saddles with the correct boundary conditions.  The first is the Hawking saddle, see
figure  \ref{fig:2wormhole}(a). The corresponding von Neumann entropy is the usual answer, $S_{\rm matter}(R(t))$, which grows linearly forever. The second is the replica wormhole, which, as we will show,  reproduces the entropy of the island rule, see figure  \ref{fig:2wormhole}(b). A replica wormhole with higher $n$ is illustrated in fig.~\ref{fig:manyreplicas}.

Replica wormholes have higher topology, so they are suppressed by factors of $e^{-S_0}$ where $S_0$ is the genus-counting parameter of JT gravity. At late times, the contribution of the Hawking saddle is heavily suppressed by the kinematics, and this is what makes it possible for the replica wormhole to take over despite the topological suppression. Indeed, the
$n^{th}$ wormhole, see fig. \ref{fig:manyreplicas}, gives a partition function $Z_n \propto 
e^{ S_0 (2 -n) } $ which leads to a $2S_0$ contribution to the  entropy.  

The wormhole topology has a saddle point at finite $n$. (We will not show this in general, but confirm it explicitly in certain limits; see below for details.) The equations that control this saddle point can be analytically continued to non-integer $n$, and used to define the replica limit $n \to 1$. 
To analyze this limit it is most convenient to assume replica symmetry and go to a quotient space which has a simpler topology but contains conical singularities and insertions of twist operators for the matter 
fields, see figure \ref{CoverThree}. In the limit $n\to 1$ both of these effects become very small and represent a small perturbation for the geometry, but they give a contribution to the entropy of precisely the same form as the gravitational generalized entropy for regions in the $n=1$ solution. The boundaries of the regions are specified by the locations of the twist operators. The replica wormholes give rise to the island contributions to the   entropy.

The physical picture that descends from accounting for these higher
topology saddles in the entropy calculation is as follows. In the
initial stages of the black hole evaporation, the quantum state of the
Hawking radiation is accurately described by quantum field theory on a
fixed background as originally studied by Hawking. This is accurate up to the Page time, defined to be the time when 
the semi-classical von Neumann
entropy of the Hawking radiation becomes equal to the the coarse-grained entropy
of the black hole. 
 At later times,  a non-perturbative effect in the gravitational path
integral results in an $O(1)$ deviation of the evolution of the
entropy of the Hawking radiation form the semi-classical result. This
is due to an exchange of dominance between the trivial topology
saddle and the wormhole saddle in the Renyi entropy calculation. This
new saddle suggests that we should think of the inside of the black hole as
a subsystem of the outgoing Hawking radiation. Namely, in the $n\to 1$ limit of the the replica trick, most of the black hole interior is included, together with the radiation, in the computation of the entropy.  This has the effect
that entanglement across the event horizon of the Hawking pairs no
longer contributes to the von Neumann entropy of the outgoing part,
while at the same time maintaining the necessary entanglement to
ensure semi-classical physics at the horizon.
  
 This paper is organized as follows. 

In section \ref{sec:CosmicStrings} we review and slightly clarify the gravitational derivation of the quantum extremal surface presciption from the replica trick in a general theory \cite{Lewkowycz:2013nqa,Faulkner:2013ana,Dong:2016hjy,Dong:2017xht}. The slight improvement is that we show that the off shell action near $n\sim 1$ becomes the generalized entropy, so that the extremality condition follows directly from the 
extremization of the action. In section 
\ref{JTplusCFT} we discuss some general aspects of replica manifolds for the case of JT gravity plus a CFT. 

In section \ref{sec:SingleInterval} we discuss the computation of the entropy for an interval that contains the degrees of freedom living at the $AdS$ boundary. In this case the quantum extremal surface is slightly outside the horizon. We set up the discussion of the Renyi entropy computations for this case. We reduce the problem to an integro-differential equation for a single function $\theta(\tau)$ that relates the physical time $\tau$ to the $AdS$ time $\theta$. We solve this equation for $n\to 1$ recovering the quantum extremal surface result. We also solve the problem for relatively high temperatures but for any $n$. 

In section \ref{sec:OneZero} we discuss the special case of the zero temperature limit, and we comment on some features of the island in that case. 

In section \ref{sec:TwoIntervals} we discuss aspects of the two intervals case, which is the one most relevant for the information problem for the eternal black hole.  

In section \ref{sec:Dictionary} we make the connection to entanglement wedge reconstruction of the black hole interior.

 We end in section \ref{sec:Conclusions} with conclusions and discussion.

\section{The replica trick for the von Neumann entropy}

\label{sec:Action}

The replica trick   for computing the von Neumann entropy is based on the observation that the computation of $Tr[\rho^n]$ can be viewed as an observable in $n$ copies of the original system \cite{Callan:1994py}. In particular, for a quantum field theory the von Neumann entropy 
of some region can be computed by considering $n$ copies of the original theory and choosing boundary conditions that connect the various copies inside the interval in a cyclic way, see e.g. \cite{Casini:2009sr} for a review.
This can be viewed as the insertion of a ``twist operator'' in the quantum field theory containing $n$ copies of the original system. 
This unnormalized correlator of twist operators can also be viewed as the partition function of the theory on a topologically non-trivial manifold, $Z_n =Z[\widetilde { \cal M}_n] = 
\langle {\cal T}_1 \cdots {\cal T}_k \rangle$. 
Then the entropy can be computed by analytically continuing in $n$  and 
setting   
\be
 S = - \left. \partial_n \left( { \log Z_n \over n} \right) \right|_{n=1} 
 \ee
 We will now review the argument for how this is computed in theories of gravity. Then we will consider the specific case of the JT gravity theory.

\subsection{The replicated action for  $n\sim 1$ becomes the generalized entropy}
 \la{sec:CosmicStrings}

In this section we review the ideas in \cite{Lewkowycz:2013nqa,Faulkner:2013ana,Dong:2016hjy,Dong:2017xht} for proving the 
holographic formula for the fine-grained entropy, or von Neumann entropy. 
We clarify why we get the generalized entropy when we evaluate the off shell 
gravity action near the $n=1$ solution.

The replica trick involves a manifold $\widetilde {\cal M}_n$ which computes
the $n^{th}$ Renyi entropy. The geometry of this manifold is completely 
fixed in the non-gravitational region, where we define the regions whose entropies we are computing\footnote{If we only had the $AdS$ theory, without an outside region, then the non-gravitational part should be viewed just as the boundary of $AdS$.}. 
  In the gravitational region we can consider any manifold, with any topology,  which obeys the appropriate boundary conditions. 
The full action for the system is a sum of the gravitational action and the partition function for the quantum fields on the geometry $\widetilde { \cal M }_n$,  
\be \la{ActOrig}
 { \log Z_n \over n}  = - { 1\over n} I_{\rm grav}[ \widetilde { \cal M}_n] + { 1 \over n}\log Z_{\rm mat}[\widetilde {\cal M}_n] \,.
 \ee
This is an effective action for the geometry and we will look for a classical solution of this combined action. In other words,   the integral over geometries is  evaluated as a saddle point. So  the metric is classical, but the equations contain the quantum expectation value of the matter stress tensor on that geometry. 
 Under the assumption of replica symmetry, we can instead consider another manifold ${\cal M}_n = \widetilde { \cal M}_n/Z_n$. This 
 manifold can be viewed as one where $n$ identical copies of the field theory are living. We have twist operators  ${\cal T}_n$ at the endpoints of the intervals in the non-gravitational region. In the gravitational region we also have twist operators ${\cal T}_n$ at the fixed points of the $Z_n$ action, where the manifold 
 ${\cal M}_n$ has conical singularities with opening angle $2\pi/n$. 
  Of course, at these points the covering manifold $\widetilde{\cal M}_n$ is smooth. 
  It is convenient to translate the problem in \nref{ActOrig} to a problem involving the manifold ${\cal M}_n$. We have $n$ copies of the matter theory propagating on this manifold. 
  In the gravitational region we can enforce the proper conical singularities in ${\cal M}_n$  by adding 
  codimension-two ``cosmic branes'' of tension 
  \be
 4 G_N  T_n =  1 - { 1 \over n}    .
 \ee 
   At these cosmic branes we also insert twist operators ${\cal T}_n$ for the $n$ copies of the 
 matter theory. In two dimensions these ``cosmic branes'' are simply points, while in four dimensions they are ``cosmic strings.'' The positions of these cosmic branes are fixed by solving the Einstein equations. We then replace the gravitational part of the action in \nref{ActOrig} by 
 \be \la{GravQuo}
 { 1 \over n } I_{\rm grav}[ \widetilde { \cal M}_n] = I_{\rm grav}[{\cal M}_n] + T_n \int_{\Sigma_{d-2} } \sqrt{g}.
 \ee
 As opposed to \cite{Lewkowycz:2013nqa}, here we add the action of these cosmic branes explicity and we also integrate
 the Einstein term through the singularity, which includes a $\delta$ function for the curvature. These two extra terms cancel out so that we get the same final answer as in \cite{Lewkowycz:2013nqa} where no contribution from the singularity was included. We will see that the present prescription is more convenient\footnote{In theories with higher
 derivatives we would need to add extra terms in the action of the cosmic brane so that they just produce a conical singularity. These presumably lead to an off shell action of the form considered in \cite{Dong:2013qoa} but we did not check this.}.
 
 In the part of the manifold where the metric is dynamical the position of these cosmic branes is fixed by the Einstein equations. Also, the reparametrization symmetry implies we cannot fix these points from the outside. 
 
 When $n=1$ we have the manifold ${\cal M}_1 = \widetilde {\cal M}_1$, 
  which is the original solution to the problem. It is a solution of the action 
 $I^{\rm tot}_1$. In order to find the manifold ${\cal M}_n$ for $n\sim 1$ we need 
 to add the cosmic branes. Then the action  is 
 \be \la{Act}
  \left( {  I^{\rm tot} \over n } \right)_{n\to  1} = I_1 + \delta \left({ I \over n} \right)
 \ee
  where $\delta I$ contains  extra terms that arise from two effects, both of which are of order $n-1$. The first comes from the tension of the cosmic brane (the second term in \nref{GravQuo}. The second comes from the insertion of the twist fields at the position of this cosmic brane. To evaluate the action perturbatively, we start from the 
 solution ${\cal M}_1$, we add the cosmic brane and twist fields, and we also consider a small deformation of
 the geometry away from   ${\cal M}_1$, where all these effects are of order $n-1$. 
 Because the ${\cal M}_1$ geometry is a solution of the original action 
 $I_1$ in \nref{Act}, any small deformation of the geometry drops out of the action. 
 For the extra term $\delta (I/n)$ in \nref{Act},  we can consider the cosmic brane action 
 and twist fields as living on the old geometry ${\cal M}_1$ since these extra terms are already 
 of order $n-1$. 
 
 Then we conclude that  the  $\delta I$ term is simply proportional to the generalized entropy 
 \be  \la{OSH}
 \delta \left( { \log Z \over n } \right)= -   \delta \left( {  I \over n } \right) = (1-n) S_{\rm gen}(w_i) = (1-n) \left[ { {\rm Area} \over 4 G_N} +S_{\rm matter} \right] ,~~~ ~~n  \sim 1
 \ee
 where we emphasized that it depends on the positions of the cosmic branes. We should emphasize that \nref{OSH} is the full off-shell action that we need to extremize  to find the classical solution of
 $I_n$ for $n \sim 1$. In this way, we obtain the quantum extremal surface prescription of \cite{Engelhardt:2014gca}, and also \cite{Ryu:2006bv,Hubeny:2007xt}.   Moreover, if we think of the cosmic strings as dynamical objects, then we can pair create them so as to form islands. This pair creation is possible in the gravity region where the tension is finite. In the region without gravity their tension is effectively infinite.

\subsection{The two dimensional JT gravity theory plus a CFT }

\la{JTplusCFT}

 \begin{figure}[t]
    \begin{center}
    \includegraphics[scale=1]{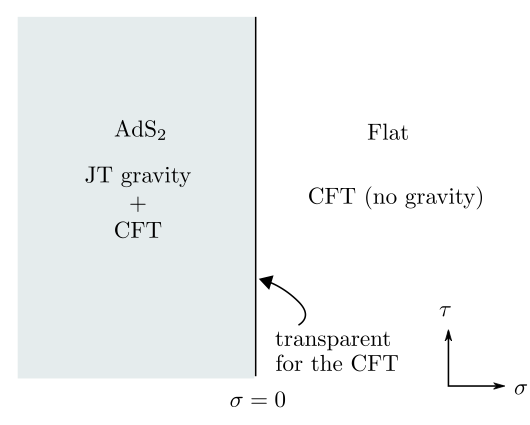}
    \end{center}
    \caption{We consider nearly-$AdS_2$ gravity with a matter CFT. The same CFT lives in an exterior flat space with no gravity. We have transparent boundary conditions for the CFT.  }
    \label{AdSandFlat}
\end{figure}

In this section we specify in more detail the theory under consideration. We have the 
Jackiw-Teitelboim gravity theory describing a nearly $AdS_2$ spacetime coupled to a matter theory that is
a CFT. In addition, we have the same  CFT  living in an  exterior flat and rigid geometry with no gravity. Since the interior and the exterior involve the same   CFT we can impose  transparent boundary conditions at the boundary,
see figure \ref{AdSandFlat}. In other words,  we have the action 
\be \la{OrigAct}
 \log Z^{\rm tot} =   { S_0 \over 4 \pi }  \left[ \int_{\Sigma_2} R +   \int_{\partial \Sigma_2} 2 K \right] + \int_{\Sigma_2} { \phi \over 4 \pi} (R+2) + { \phi_b \over 4 \pi}  \int_{\partial \Sigma_2} 2 K    + \log Z_{CFT}[g]
 \ee
 where the CFT action is defined over a geometry which is rigid in the exterior region and is dynamical in the interior region.   
 We are setting $4G_N=1$ so that the area terms in the entropies will be just given by the value of $\phi$, ${{\rm Area } \over 4 G_N} = S_0 + \phi$. 
 
\begin{figure}[t]
    \begin{center}
    \includegraphics[scale=0.65]{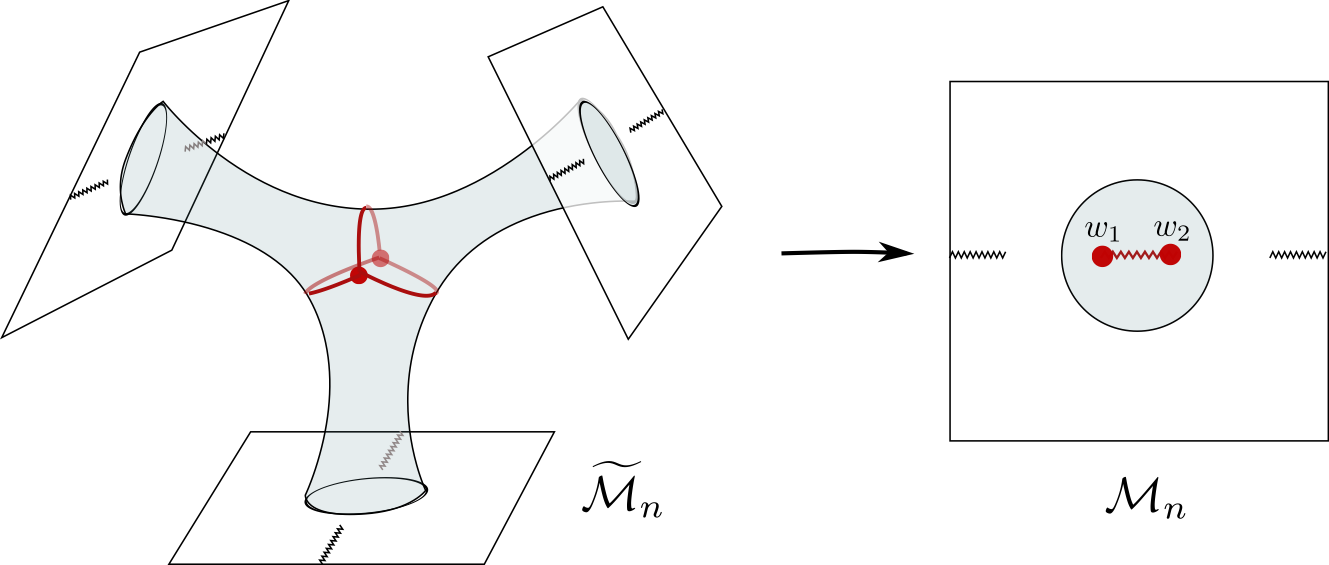}
    \end{center}
    \caption{Here we display the replica manifold, $\widetilde {\cal M}_3$, 
    and also the manifold ${\cal M}_3 = \widetilde {\cal M}_3/Z_3$ which has the topology of the disk with conical singularities at two points $w_1$ and $w_2$ which corresponds to the fixed points of the $Z_3$ action on $\widetilde {\cal M}_3$. We parametrize this disk in terms of the holomorphic coordinate $w$. The exterior regions of $\widetilde{\cal M}_n$ are also glued together cyclically along the cuts. }
    \label{CoverThree}
\end{figure}

 In this theory,  we want to consider the replica manifolds described above, see figure \ref{fig:manyreplicas}. Because we consider replica symmetric solutions, it is convenient to quotient by $Z_n$ and discuss a single manifold with $n$ copies of the matter theory on it. 
 In other words, we go from the action \nref{OrigAct} on $\widetilde {\cal M}_n$ to a problem 
 on ${\cal M}_n = \widetilde {\cal M}_n/Z_n $. We find that this simplifies a bit the description of the manifold, see figure \ref{CoverThree}. Namely, the manifold ${\cal M}_n$  can be viewed as a disk with conical singularities and with twist operators for the matter theory inserted at these singularities. These are the cosmic branes discussed in section \ref{sec:CosmicStrings}. The final gravitational action is as in  \nref{OrigAct} but 
  with an additional factor of $n$ and extra terms that produce the conical singularities 
  \be \la{newgra}
  -{ 1 \over n} I_{\rm grav} =  
{  S_0 \over 4 \pi }  \left[ \int_{\Sigma_2} R +   \int_{\partial \Sigma_2} 2 K \right] +  \int_{\Sigma_2} { \phi \over 4 \pi} (R+2) +  { \phi_b \over 4 \pi}  \int_{\partial \Sigma_2} 2 K  -  (1-{1 \over n}) \sum_i [ S_0 + \phi(w_i) ]
\ee
where $w_i$ are the positions of the conical singularities, or cosmic branes (which are just instantons  or -1 branes). 
We can consider \nref{newgra} as a new gravity theory and add $n$ copies of the CFT. In addition, 
we put twist fields at the positions $w_i$ of the cosmic branes.  It might look like we are breaking reparametrization invariance when when add these terms.  Reparametrization symmetry is restored because $w_i$ are dynamical variables which can be anywhere on the manifold and will be fixed by the equations of motion.

 We treat the CFT as a quantum theory and evaluate its partition function. Then we solve the classical equations for the metric and dilaton inserting the quantum expectation value of the stress tensor. This approximation is particularly appropriate when the central charge is large $c\gg 1$. 
 So we imagine that we are in that regime for the simple euclidean solutions we discuss here. The approximation can also be justified in other regimes where  the entanglement entropy of matter is large for kinematical reasons.  
 However, this description is {\it not} correct when we need to include the quantum aspects of gravity. That computation should be done in the original manifold and the fact that the fluctuations can break the replica symmetry is important. 
 
 We can define an interior 
  complex coordinate $w$ where the metric for the manifold ${\cal M}_n$ in the gravitational region is  
  \be
   ds^2 = e^{ 2 \rho } d w d \bar w   ~,~~~~~{\rm with} ~~ |w| \leq 1\,. \la{MetrDis}
   \ee
   The boundary of $AdS_2$ is at $|w|=1$, or $w = e^{ i \theta} $.   
   \eqref{MetrDis} is a constant curvature metric on the disk $|w|\leq 1$ with conical singularities
   at certain values $w_i$ with opening angle $2\pi/n$. This type of metric is enforced by the dilaton equation of motion in \nref{newgra} 
  \be \la{RhoEq}
 - 4 \partial_w \partial_{\bar w} \rho +
    e^{  2 \rho} = 2 \pi (1 - { 1 \over n } ) \sum_i \delta^2(w -w_i) 
   \ee
   On this space we have $n$ copies of 
   the CFT and we have twist fields inserted at the conical singularities. 
   Notice that once we impose this equation,  the contributions in    \nref{newgra}  from the delta functions in the curvature cancel against the explicit cosmic brane action terms, as we anticipated in section \ref{sec:CosmicStrings}.

\begin{figure}[t]
    \begin{center}
    \includegraphics[scale=1]{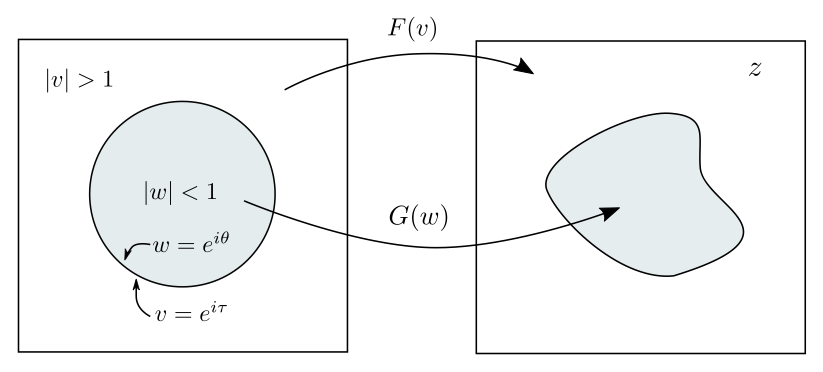}
    \end{center}
    \caption{ The conformal welding problem.  We are given two disks, one parametrized by
     $|w|\leq 1$ and another by
    $|v| \geq 1$ with their boundaries glued in terms of a given function $\theta(\tau)$ where
    $w =e^{i\theta}$ and $v=e^{i\tau}$. Then we need to find holomorphic maps of each disk to a region of the complex $z$ plane so that they are compatible at the boundary. The functions $F$ and $G$ are only required to be holomorphic inside their respective disks.   }
    \label{WandZplanes}
\end{figure}

   This metric should be joined to the flat space outside.
    We consider a finite temperature 
   configuration where $\tau \sim \tau + 2 \pi$. For general temperatures,  all we need to do is to rescale $\phi_r \to  2 \pi \phi_r/\beta $. In other words,
    the only dimensionful scale is $\phi_r$, so the only dependence on the temperature for dimensionless quantities   is through $\phi_r/\beta$.   
   We   define the coordinate $v = e^y$. So the physical half cylinder $\sigma \geq 0$ corresponds to $|v|\geq 1$. 
   At the boundary we have that $w = e^{ i \theta(\tau)}$, $v = e^{ i \tau }$. 
   Unfortunately,  we cannot extend this to a holomorphic map in the interior of the disk.
    However, we can find another coordinate $z$ such that there are holomorphic maps from $|w|\leq 1$ and $|v|\geq 1$ to the coordinate $z$, see figure \ref{WandZplanes}. 
   
   In other words, it is possible to find two functions $G$ and $F $ such that 
   \bea \la{WeldEqns}
    z &=& G(w) ~,~~~{\rm for}~~~|w|\leq 1 ~
    \cr  ~~~~~~ z &=& F(v) ~,~~~~{\rm for}~~~|v|\geq  1 
   \cr
   &~&  G(e^{ i\theta(\tau)}) = F(e^{i\tau} ) ~,~~~{\rm for} ~~~|w|=|v|=1\,.
   \eea
   The functions $F$ and $G$ are holomorphic in their respective domains (they do not have to be holomorphic at the boundary). 
   The problem of finding $F$ and $G$ given $\theta(\tau)$ is called the 
   ``conformal welding problem,'' see \cite{Mumford} for a nice discussion.\footnote{We thank 
   L. Iliesiu and Z. Yang for discussions on this problem, and A. Lupsasca for pointing out the connection to \cite{Mumford}.}  $F$ and $G$   end up depending non-locally on $\theta(\tau)$ and they
     map the inside and outside disks to the inside and outside of some irregular region in the complex plane, see figure \ref{WandZplanes}. In our problem, $\theta(\tau)$ arises as the reparametrization mode, or ``boundary 
     graviton'' of the nearly-$AdS_2$ gravity theory \cite{Jensen:2016pah,Maldacena:2016upp,Engelsoy:2016xyb}.
   
   When $n=1$, we have a trivial stress tensor in the $z$ plane. We then insert the twist operators in the outside region, and also in the inside region. We are free to insert as many
    conical singularities and twist fields in the inside as we want. This amounts to considering various numbers of islands in the gravity region. We will only discuss cases with one or two inside insertions in the subsequent sections. 
     This gives us a non-trivial stress tensor $T_{zz}(z)$ and $T_{\bar z \bar z}(\bar z)$. We can then compute the physical stress tensor that will appear in the equation of motion using the 
   conformal anomaly, 
   \be \la{StressT}
    T_{yy} = \left( { d F(e^{ i y} ) \over dy } \right)^2 T_{zz} - 
    { c \over 24 \pi } \{ F(e^{i y}), y \} 
    \ee
    and a similar expression for $T_{\bar y \bar y } $. The expression for the physical stress tensor in the $w$ plane  involves the function $G$ and also a conformal anomaly contribution from $\rho$ in the metric \nref{MetrDis}.
    
  Let us now turn to the problem of writing the equations of motion for the boundary reparametrization mode.  
  Naively we are tempted to write the action just as    $\{ e^{ i\theta } , \tau \}$. This would be correct if there were no conical singularities in the interior. However, the presence of those conical singularities implies that the metric \nref{MetrDis} has small deviations
    compared to the metric of a standard hyperbolic disk 
    \be \la{Metdr}
    ds^2 = e^{ 2 \rho} dw d\bar w ~,~~~~~~~ e^{ 2 \rho} = { 4 \over (1 - |w|^2)^2 } e^{ 2 \delta \rho } 
    \ee
    where $\delta \rho $ goes as 
    \be \la{DelRhoe}
     \delta \rho  \sim -{ (1-|w|)^2 \over 3 } U(\theta)  ~,~~~~~{\rm as}~~~|w| \to 1\,.
     \ee 
   The function $U$ depends on the positions of the conical singularities and therefore also on
   the moduli of the Riemann surface. 
   This then implies that the Schwarzian term, and the full equation of motion can now be written as 
   \be \la{EOMFin}
   { \phi_r \over 2 \pi } { d \over d \tau } \left[ \{ e^{ i \theta} , \tau \} + U(\theta) {\theta'}^2  \right] = i (T_{yy} - T_{\bar y \bar y } ) = T_{\tau \sigma}\,.
   \ee
   The term in brackets is proportional to the energy.  This equation relates the change in energy  to the energy flux from the flat space region. Here the flux of energy on the right hand side is that of one copy, or the flux of the $n$ copies divided by $n$. 
   The action can be derived from the extrinsic curvature term in the same way that was discussed in 
  \cite{Jensen:2016pah,Maldacena:2016upp,Engelsoy:2016xyb}, see appendix \ref{GravAct}, where we also discuss the 
  explicit derivation of the equation of motion \nref{EOMFin}. 
   
   There are also equations that result from varying the moduli of the Riemann surface, or the 
   positions of the conical singularities. They  have the form 
   \be \la{PartSing}
    -(1 - { 1 \over n}) \partial_{w} \phi(w_i) + \partial_{w_i } \left( { \log Z^{\rm mat}_n \over n } \right) =0 \,,
   \ee 
   where we used that the $w_i$ dependence of the gravitational part of the action comes only from the last term in \nref{newgra}.   
  
  In the $n\to 1$ limit we can replace the $n=1$ value for the dilaton in 
   \nref{PartSing}. Similarly the value of $\log Z^{\rm mat}_n/n$ near $n=1$ involves the matter entropy. 
   Therefore \nref{PartSing}  reduces to the condition on the extremization of the generalized 
   entropy, as we discussed in general above. 
   
   For general $n$, we need to compute the dilaton by solving its equations of motion 
   in order to write  \nref{PartSing}. This can be done using the expression for the stress tensor in the interior of the disk.   
    We have not attempted to simplify it further. However, we should note that for the particular case of one interval, discussed in section \ref{sec:SingleInterval}, there is only one point and there are no moduli for the Riemann surface. Therefore this equation is redundant and in fact, it is contained in \nref{EOMFin} as will be discussed in section \ref{sec:SingleInterval}. 
  
 Next we apply this general discussion to the calculation of the entropy of various subregions of the flat space CFT. The goal is to understand how configurations of the gravity region contribute to the entropy of those CFT regions.

\newcommand{\by}{\bar{y}}
\newcommand{\bw}{\bar{w}}
\newcommand{\bx}{\bar{x}}
\newcommand{\bu}{\bar{u}}
\newcommand{\bv}{\bar{v}}
\renewcommand{\ba}{\bar{a}}
\newcommand{\E}{{\cal E}}
\newcommand{\Hilbert}{{\rm H}}

\section{Single interval at finite temperature}

\la{sec:SingleInterval}

\begin{figure}
\begin{center}
\includegraphics[scale=0.5]{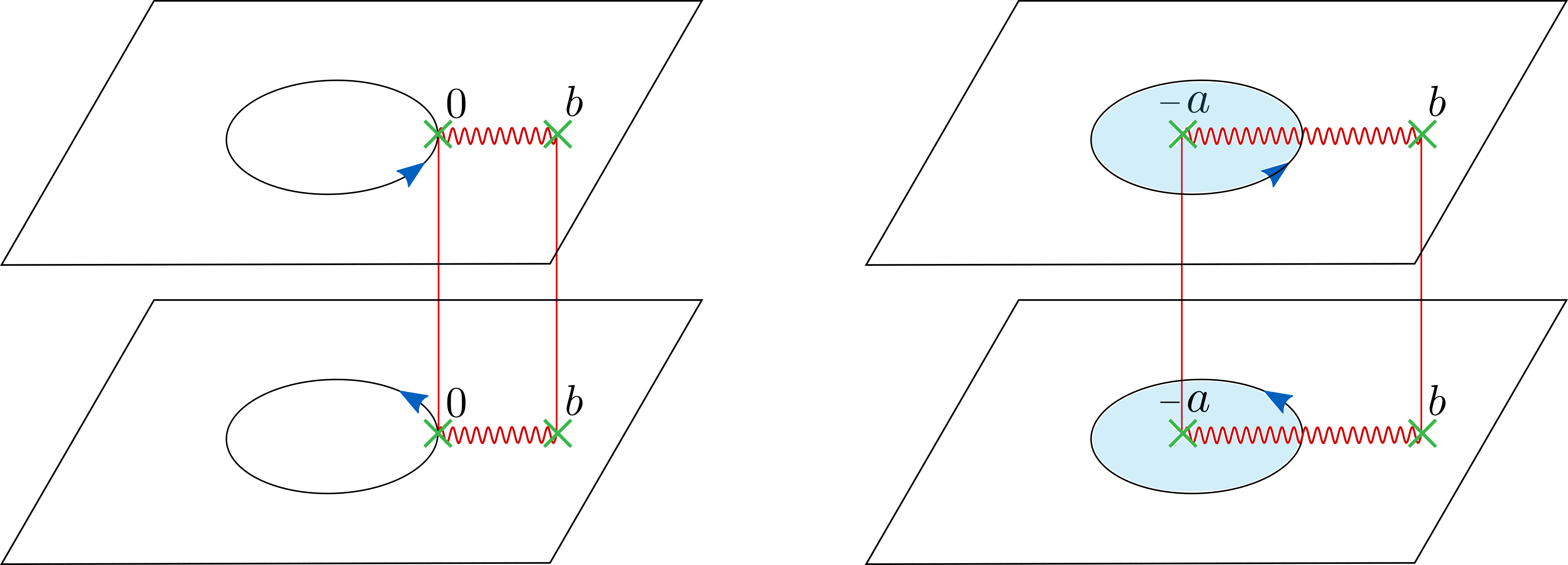}

(a) ~~~~~~~~~~~~~~~~~~~~~~~~~~~~~~~~~~~~~~~~~~~~~~~~~~~~ ~~~~~(b)
\end{center}
\caption{\small  
(a) We have a flat space field theory on the exterior of the disk. The disk is hollow in this picture, and will be filled in with gravitational configurations subject to the boundary conditions on the unit circle. 
 This boundary is connected into a single long circle $n$ times longer than the original one. This is indicated by the blue arrow which tells you how to go around the cut. (b) The disk is filled in with a gravitational configuration with the topology of a disk which ends on the elongated unit circle.  This configuration can be represented by adding a branch point inside. Note that the local geometry at the branch point ``$-a$'' is completely smooth.  
\label{SingleIntervalSheets}}
\end{figure}

 We begin with the simple case of a single interval that contains one of the AdS$_2$ boundaries, as shown in figure 
 \ref{SingleIntervalSheets}(a). This is the interval $B \equiv  { [ 0, b ]}$.

To compute the entropy of this region we must consider the Euclidean path integral that evaluates the trace of powers of the density matrix $\Tr[\rho_B^n]$. This is given by the path integral on $n$ copies of the theory identified across the region $B$, as shown in figure \ref{SingleIntervalSheets}. The crucial point is that the presence of the branch point on the unit circle, which is where the asymptotic AdS boundary lives, elongates this circle by a factor of $n$. The Euclidean gravity configurations we must consider are all smooth manifolds with a single boundary that is identified with this elongated AdS boundary.

The simplest configuration to consider will be that with the topology of a disk. All other higher genus manifolds will be subleading since each extra handle will come with a cost of $e^{-S_0}$. Filling out the gravity region has the effect of extending the identification across the different sheets into the gravity region, which ends on some point ``$-a$'' in figure \ref{SingleIntervalSheets}. The location of the point ``$-a$'' will be dynamically determined by the saddle point of the path integral.

\begin{figure}
\begin{center}
\includegraphics[scale=0.7]{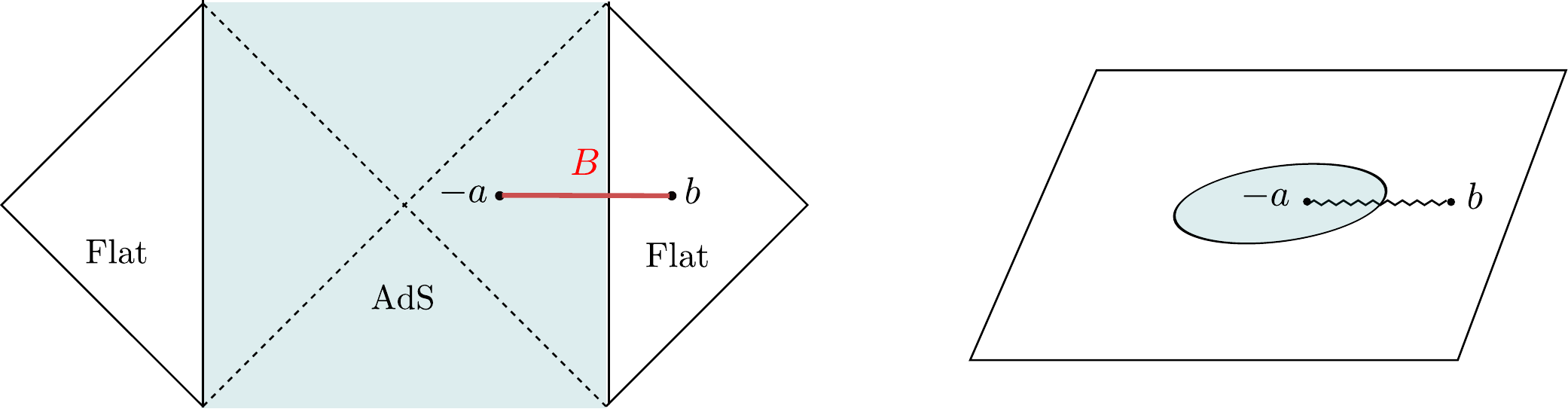}
\end{center}
\caption{\small The single interval configuration in Lorentzian signature (left) and in Euclidean signature (right).  
\label{SingleLandE}}
\end{figure}

We will now construct replica wormholes explicitly for a single interval in the eternal black hole in AdS$_2$.  The Lorentzian and Euclidean geometries are shown in figure \ref{SingleLandE}. 
 We will first review of the result of the QES calculation \cite{Almheiri:2019yqk}, then proceed to derive it from replica wormholes.

\subsection{Geometry of the black hole}

The metric of eternal  black hole, glued to flat space on both sides, is
\be\label{dsyin}
ds^2_{\rm in} = \frac{4\pi^2}{\beta^2}\frac{dy d\by}{ \sinh^2 \frac{\pi}{\beta}(y+\by)} , \qquad
ds^2_{\rm out} = \frac{1}{\epsilon^2} dy d\by \ , 
\ee
\be
y = \sigma + i \tau ,\qquad \by = \sigma - i \tau  ~,~~~~~\tau = \tau + \beta  \ .
\ee
The subscript `in' refers to the gravity zone, and `out' refers to the matter zone\footnote{The Poincare coordinates are $ x = \tanh{ \pi y \over \beta } $, $ds^2_{\rm in} = 4 d x d\bar x/(x+\bar x)^2$. The  Schwarzschild coordinates are \be
y = \frac{\beta}{2\pi}\log \frac{r}{\sqrt{r(r + 4\pi/\beta)}}  + i \tau  \ ,  \qquad
ds^2_{in} = r(r + \frac{4\pi}{\beta})d\tau^2 + \frac{dr^2}{r(r + \frac{4\pi}{\beta})}  \ .
\ee   }. The interface is along the circle $\sigma = -\epsilon$. 
 Lorenztian time $t$ is  $\tau = -it$. 
The welding maps of figure \ref{WandZplanes} are trivial  and we have
\begin{align} \la{wexpy}
z=v= w = e^{2 \pi y /\beta} \ , \qquad y = \frac{\beta}{2\pi}{\log w} \ .
\end{align}
The Euclidean solution is   therefore the $w$-plane with gravity inside the unit disk, $|w|  <  1 - \frac{2\pi \epsilon}{\beta}$. The metric is
\be\label{xmet}
ds^2_{\rm in} =  \frac{4dw d\bw}{(1-|w|^2)^2} , \qquad   \qquad 
ds^2_{\rm out} =\frac{\beta^2}{4\pi^2\epsilon^2} \frac{dw d\bw}{|w|^2} \ .
\ee
The dilaton, which is defined only on the inside region, is rotationally invariant on the $w$-plane, 
\be \la{DilExpr}
\phi = \frac{2\pi \phi_r}{\beta} \frac{1 + |w|^2}{1 - |w|^2} = -{ 2 \pi \phi_r \over \beta }{ 1 \over \tanh{ 2\pi \sigma \over \beta } }   \ .
\ee
 with $\phi = \phi_r/\epsilon$ at the boundary. In what follows, we will usually set $\epsilon = 0$, and rescale the exterior coordinate by $\epsilon$ so that $ds^2_{out} = dy d\by$. 

\subsection{Quantum extremal surface}

We now review the computation of the entropy of the region $B = { [0,b]}$ which includes the $AdS_2$ 
boundary, see figure \ref{SingleIntervalSheets}.  
In gravity this will involve an interval $[-a,b]$, with  
 $a, b > 0$, see figure \ref{SingleLandE}.

The generalized entropy of the region $[-a,b]$ is
\be \la{SgenGen}
S_{\rm gen} = S_0 + \phi(-a) +  
S_{\rm CFT}([-a,b]) \ .
\ee
The entanglement entropy of a CFT on the interval $[w_1,w_2]$ in the metric $ds^2=\Omega^{-2} dw d\bw$ is
\be\label{scftgeneral}
S_{\rm CFT}(w_1, w_2) = \frac{c}{6}\log \left(\frac{ |w_1 - w_2|^2 }{\epsilon_{1,UV} \epsilon_{2,UV} \Omega(w_1, \bw_1) \Omega(w_2, \bw_2) } \right) \ .
\ee
Using the map $w = e^{2 \pi y/\beta}$ and the conformal factors in \eqref{xmet} this becomes
\begin{align}
S_{\rm CFT}([-a,b]) &= 
 \frac{c}{6}\log \left(  \frac{2\beta\sinh^2 \left( \frac{\pi}{\beta} (a+b)\right) }{\epsilon_{a,UV} \epsilon_{b,UV} \pi    \sinh \left( \frac{2\pi a}{\beta} \right)}\right)
\end{align}
Then, using the dilaton in \nref{DilExpr}, \nref{SgenGen} becomes 
\begin{align}\label{sgens}
S_{\rm gen}([-a,b]) &=
S_0 + \frac{2\pi \phi_r}{\beta}\frac{1}{\tanh \left( \frac{2\pi a}{\beta} \right)  } 
+   \frac{c}{6}\log \left( \frac{2\beta\sinh^2\left( \frac{\pi}{\beta}(a+b) \right) }{ \pi \epsilon \sinh \left( \frac{2 \pi a}{\beta}\right)} \right) \ .
\end{align}
The UV divergence $\epsilon_{a,UV}$ was absorbed into $S_0$ and we dropped the outside one at point $b$. 
The quantum extremal surface is defined by extremizing $S_{\rm gen}$ over $a$ 
\begin{align} \label{tzeroex2}
\partial_a S_{\rm gen} =0 ~~\to~~~~~~ \sinh \left(\frac{2\pi a}{\beta} \right) =  \frac{12 \pi \phi_r}{\beta c} \frac{\sinh \left( \frac{\pi}{\beta}(b+a)\right)}{\sinh \left( \frac{\pi}{\beta}(a-b)\right)}
\end{align}
This is a cubic equation for $e^{2\pi a/\beta}$. For $b \gtrsim \frac{\beta}{2\pi}$  and $\phi_r/(\beta c) \gtrsim 1$,  the solution is
\be \la{smallAB}
a \approx b  + \frac{\beta}{2\pi}\log \left( \frac{24 \pi \phi_r}{\beta c}\right) ~, ~~~~~~{\rm or } ~~~~~ e^{ - { 2\pi a \over \beta} }\approx  { \beta c \over 24 \pi \phi_r } { e^{-{ 2 \pi b \over \beta } } 
} 
\ee
Since we've restricted to one side of the black hole in this calculation, the configuration is invariant under translations in the Schwarzschild $t$ direction. Therefore the general extremal surface at $t\neq 0$ is related by a time translation; for an interval that starts at $t_b$ and $\sigma_b = b$, the other endpoint is at 
  $t_a = t_b$ and $\sigma_a=-a$, with $a$ as in   \eqref{tzeroex2}.

\subsection{Setting up the  replica geometries }

We will do the replica calculation in Euclidean signature, with $a,b$ real. We set $\beta = 2\pi$, and reintroduce it later by dimensional analysis. 

The replica wormhole that we seek is an $n$-fold cover of the Euclidean black hole, branched at the points $a$ and $b$, 
see figure \nref{SingleLandE}. 
This manifold will have a nontrivial gluing at the unit circle (unlike the black hole itself), so it is more convenient to introduce different coordinates on the inside and outside. We use $w$, with $|w| < 1$, for the inside and 
$v = e^y$, with $|v|>1$ for the outside.    The gluing function is $\theta(\tau)$, with 
$
w = e^{i \theta} $, $ v = e^{i\tau} $, as in \nref{WeldEqns}. 
We write the branch points as 
\be
w = A = e^{-a} \ , \qquad v = B = e^b \ .
\ee

The Schwarzian equation is simplest in a different coordinate, 
\be \la{wtildew}
\widetilde{w} = \left( \frac{w - A}{1 - A w} \right)^{1/n} \ .
\ee
This coordinate uniformizes $n$ copies of the unit disk, so here we have the standard hyperbolic metric,
\be \la{metrictil}
ds^2_{in} = \frac{4 |d\tilde{w}|^2}{(1 - |\tilde{w}|^2)^2} \ .
\ee
Defining   $\tilde w = e^{ i \tilde \theta}$ at the boundary, the Schwarzian equation is
\be\label{schwtilde}
\frac{\phi_r}{2\pi} \p_{\tau} \{e^{ i \tilde \theta}, \tau\} = i(T_{yy}(i\tau) - T_{\by\by}(-i\tau)) \ .
\ee
We can now return to the $w$-disk using the Schwarzian composition identity
\be
\{e^{ i \tilde \theta} , \tau\} = \{ e^{i \theta}, \tau\}  + \frac{1}{2}\left(1- \frac{1}{n^2} \right) R(\theta) \ , 
\ee
with
\be \la{Rdefi}
R(\theta) = - \frac{(1-A^2)^2 ( \partial_\tau \theta)^2 }{ |1  - Ae^{i\theta}|^4  } \ .
\ee
This puts the equation of motion \eqref{schwtilde} into exactly the form of equation  \eqref{EOMFin}, which we have just derived by a slightly different route. In appendix \ref{GravAct} we show that they are equivalent. 

The stress tensor appearing on the right-hand side of \eqref{schwtilde} is obtained through the conformal welding. That is, we define the $z$ coordinate by the map $G$ on the inside and $F$ on the outside as in \eqref{WeldEqns}. These maps each have an ambiguity under $SL(2,C)$ transformations of $z$, which we may use to map the twist operator at $w = A$ to $z=0$, and the twist operator at $v = B$ to $z = \infty$. We further discuss the symmetries of the conformal welding problem in appendix \ref{app:welding}.

The $z$-coordinate covers the full plane holomorphically. It has twist points at the origin and at infinity, which can be removed by the standard mapping, $\tilde{z} = z^{1/n}$. On the $\tilde{z}$ plane, the stress tensor vanishes, so on the $z$-plane,
\be
T_{zz}(z) = -\frac{c}{24\pi} \{ z^{1/n}, z \} = -\frac{c}{48\pi}\left(1 - \frac{1}{n^2}\right) \frac{1}{z^2} \ .
\ee
Finally the stress tensor $T_{yy}$ comes from inverting the conformal welding map to return to the $v$-plane, and using $v = e^{y}$:
\be
T_{yy}(y) = e^{2y}\left[ F'(v)^2 T_{zz} - \frac{c}{24\pi}\{ F, v \} \right]  - \half \ .
\ee
Putting it all together, the equation of motion \eqref{schwtilde} is
\be \label{singleintervalEOM}
\frac{24 \pi \phi_r  }{c \beta}\p_\tau
\left[ \{ e^{i\theta(\tau)}, \tau \} + \frac{1}{2} (1- \frac{1}{n^2} ) R(\theta(\tau)) \right] = 
ie^{2i\tau}\left[ 
 - \frac{1}{2}(1 - \frac{1}{n^2}) \frac{F'(e^{i\tau})^2}{F(e^{i\tau})^2} -  \{F, e^{i\tau} \} \right]  + cc
\ee
This equation originated on the smooth replica manifold $\widetilde{{\cal M}}_n$, but has now been written entirely on the quotient manifold ${\cal M}_{n} = \widetilde{{\cal M}}_n/\mathbf{Z}_n$.  We have restored the nontrivial temperature dependence\footnote{The trivial temperature dependence is restored by $\tau \to {  2 \pi \over \beta }\tau_{phys}$, with $\tau_{phys}$ the physical Euclidean time with period $\beta$.}. In particular, note that $\theta(\tau+ 2\pi) = \theta + 2 \pi$. The $\tau \rightarrow -\tau$ symmetry of the insertions allows us to choose a function $\theta(\tau) = - \theta(-\tau)$ which will automatically obey $\theta(0)=0$, $\theta''(0)=0$. In addition, we should then impose $\theta(\pi) = \pi$ and $\theta''(\pi) =0$. The problem now is such that $n$ appears as a continuous parameter and there is no difficulty in analytically continuing in $n$. 

This is our final answer for the equation of motion at finite $n$. It is quite complicated, because the welding map $F$ depends implicitly on the gluing function $\theta(\tau)$.  We will solve it in two limits: $\beta \to 0$ at any $n$, and $n \to 1$ at any $\beta$.

\subsection{Replica solution as $n \to 1$}

We will now show that the equation of motion \eqref{singleintervalEOM} reproduces the equation for the quantum extremal surface.

We start with the solution for $n=1$. In this case the welding problem is trivial and we can set $w =v $ everywhere. 
It is convenient to set 
\be \la{zChoice}
 z = F(v) = { v - A \over B - v} = G(w) ~,~~~~~w=v
 \ee
   At $n=1$ any choice of $A$ can do. Different choices of $A$ can be related by an 
 $SL(2,R)$ transformation that acts on $w$. It will be convenient for us to choose $A$ so that when we go to $n\sim 1$, it corresponds to the position of the conical singularity.

We now go near $n \sim 1$ and expand
\be
e^{i\theta} = e^{i\tau} + 
e^{ i \tau} i \delta \theta(\tau) \ ,
\ee
where $\delta \theta$ is of order $n-1$. 
We aim to solve
 \eqref{singleintervalEOM} for $\delta \theta$. 
The first step is to find the welding map perturbatively in $(n-1)$. 
In appendix \ref{app:welding}, we show that
\be
e^{2i\tau} \{ F , e^{i \tau} \} = -\delta\{e^{i\theta} , \tau \}_-  = -(\delta \theta''' + \delta \theta' )_-
\ee
where we used
\be \delta \{ e^{i\theta} ,\tau \} \equiv \{ e^{ i \tau + i \delta \theta} , \tau \} - \{ e^{ i \tau } , \tau \} = \delta \theta''' + \delta \theta' 
\ee
The minus subscript indicates that this is projected onto negative-frequency modes. This can be written neatly using the Hilbert transform, $\Hilbert$, which is defined by the action $\Hilbert \cdot e^{i m \tau} = - \mbox{sgn} (m) e^{im \tau}$ (and $\Hilbert \cdot 1 = 0$). Then
\be \la{HilH}
e^{ 2 i \tau } \{ F , e^{i \tau} \} = -\half (1 +\Hilbert)
(\delta \theta''' + \delta \theta' )  .
\ee
Wherever else $F$ appears in \eqref{singleintervalEOM}, it is multiplied by $(n-1)$, so there we can set $F = \frac{v-A}{B-v}$, as in \nref{zChoice}. Therefore the equation of motion for the perturbation is
\be\label{hilbertp}
\p_\tau (\delta \theta''' + \delta \theta' ) + \frac{ic}{12 \phi_r} \Hilbert \cdot (\delta \theta''' + \delta \theta' ) = (n-1) \left[ \frac{c}{12\phi_r}{\cal F} -\p_\tau R(\tau) \right]
\ee
where 
\be\label{fluxdefFull}
{\cal F} = -i  \frac{e^{2 i \tau}  (A - B)^2}{ (e^{i \tau} - A )^2 (e^{ i \tau} - B)^2} + cc  \ .
\ee
Equation \eqref{hilbertp} is nonlocal, due to the Hilbert transform. 
We can solve it by expanding both sides in a Fourier series. 
The important observation is that, due to the structure of derivatives in each term of the left hand side of \nref{hilbertp},  the terms with 
Fourier modes of the form $e^{ik \tau}$ for $k = 0, \pm 1$ are automatically  zero in the left hand side. Therefore, in order to solve this equation, we must impose the same condition on the right-hand side. The $k =  1$ mode requires
\be
\int_{0}^{2\pi}d\tau e^{-i\tau} \left( \frac{c}{12\phi_r}{\cal F} -\p_\tau R(\tau) \right) = 0 \ .
\ee
Doing the integrals, this gives the condition
\be
\frac{c}{6 \phi_r} \frac{ \sinh \frac{a-b}{2} }{ \sinh \frac{b+a}{2} }  = \frac{1}{\sinh a } ~.~~~~~~    \ee
This matches the equation for the quantum extremal surface \eqref{tzeroex2} that came from the derivative of the generalized entropy.
The term with $k=0$ is automatically zero in the right hand side, as  $\partial_\tau R$ is explicitly a total derivative and   $ \int_0^{2\pi } d\tau {\cal F} = 0$. 

Thus we have reproduced the QES directly from the equations of motion. Once the QES condition is imposed, it is straightforward to solve  for the rest of the 
the Fourier modes of $\delta \theta$ to confirm that there is indeed a solution.

The Hilbert transform that appeared in the equations of motion \nref{hilbertp} has a natural interpretation in Lorentzian signature as  the term responsible for dissipation of an evaporating black hole into Hawking radiation. This is elaborated upon in appendix \ref{app:lorentzian}.

\subsection{Entropy}

To calculate the entropy, we must evaluate the action to leading order in $n-1$. By the general arguments of section \ref{sec:CosmicStrings}, this will reproduce the generalized entropy in the bulk. Here we will check this explicitly.

The gravitational action \eqref{newgra} in terms of the Schwarzian is 
\be
-I_{\rm grav} =  S_0 + \frac{\phi_r}{2\pi} n \int_0^{2\pi} d\tau\left( \{e^{i\theta}, \tau\}  + \frac{1}{2}(1-\frac{1}{n^2}) R(\theta) \right) \ .
\ee
The first term is $-S_0$ times the Euler characteristic of the replica wormholes, $\chi = 1$ in this case.
After normalizing, the contribution to $-\log \Tr (\rho_R)^n$ for $n \approx 1$ is
\be\label{igrav1}
-I_{\rm grav}(n) + n I_{\rm grav}(1) \approx (1-n)S_0  + (n-1) \frac{\phi_r}{2\pi} \int_0^{2\pi} d\tau R(\tau) +(n-1)\frac{\phi_r}{2\pi}  \int_0^{2\pi}d\tau \p_n \{e^{i\theta}, \tau\}  \ .
\ee
The first two terms give the area term in the generalized entropy. The second term is the dilaton at the branch point,
\be
\frac{\phi_r}{2\pi} \int_0^{2\pi} d\tau R(\tau)  =  - { \phi_r \over \tanh a } 
\ee

The leading term in the matter action is the von Neumann entropy of the CFT,\footnote{This is derived in the standard way, for example by integrating the CFT Ward identity for $\p_b \log Z_M$ \cite{Calabrese:2004eu}.}  plus a contribution from an order $(n-1)$  change in the metric  
\be \label{imat1}
\log Z^{\rm mat}_{n}  - n \log Z^{\rm mat}_1 = -(n-1) S_{\rm bulk}([-a,b])  + \delta_g \log Z_M \ .
\ee
The matter action is evaluating on the manifold with the dynamical twist point in the gravity region, so the bulk entropy includes the island, $I$. By the equation of motion at $n=1$, the last term in \eqref{igrav1} cancels the last term in \eqref{imat1}, leading to
\be
\log \Tr (\rho_B)^n \approx (1-n) S_{\rm gen}([-a,b]) ~~\to ~~S( { [0,b] } ) = S_{\rm gen} ( [-a,b]) \ ,
\ee
as predicted by the general arguments reviewed in section \ref{sec:CosmicStrings} \cite{Dong:2017xht}.

\subsection{High-temperature limit}

For general $n$ is is convenient to write the equation as follows. 
The problem has an $SL(2,R)$ gauge symmetry that acts on $w$ and $A$. We can use it to gauge fix $A=0$. Then the equation \nref{singleintervalEOM} becomes
\be \label{EOMfixed}
\p_\tau
\{ e^{i\theta(\tau)/n}, \tau \}  = \kappa 
ie^{2i\tau}\left[ 
 - \frac{1}{2}(1 - \frac{1}{n^2}) \frac{F'(e^{i\tau})^2}{F(e^{i\tau})^2} -  \{F, e^{i\tau} \} \right]  + cc
\ee
Where we introduced
\be
\kappa \equiv  \frac{c \beta}{24 \pi \phi_r} 
\ee
This is proportional to the ratio of $c$ and the near extremal entropy of the black hole $S-S_0$. 
When this parameter is small, the equations simplify. This essentially corresponds to weak gravitational coupling. In this section we will study the equations for small $\kappa \ll 1$.

To leading order, we can ignore the effects of welding and set $F=G$ with
\be\label{Fleading}
F(v) =  \frac{v}{B-v}  \ , ~~~~~~~~~~~ G(w) =  \frac{w}{B-w}
\ee
This eliminates all the effects of welding, so the equation of motion is a completely explicit differential equation for $\theta(\tau)$.
We expand
\be
\theta(\tau) = \tau +  \delta \theta(\tau) \ ,
\ee
with $\delta \theta$ of order $\kappa$. 
The equation \eqref{EOMfixed} is
\be\label{weakeom}
\p_{\tau} \left( \delta \theta''' + \frac{1}{n^2} \delta\theta' \right)   = \frac{\kappa}{2}(1 - \frac{1}{n^2}){\cal F} 
\ee
with
\be\label{fluxdef}
{\cal F} = -i    \left(1 - { e^{ i \tau} \over  B}\right)^{-2} + cc \ .
\ee
We can expand this in a power series. The constant Fourier mode is absent in the right hand side of \nref{weakeom}. After solving \nref{weakeom} in Fourier space we get 
\be \la{thetF}
\delta \theta = - i { \kappa \over 2 } ( 1 - { 1 \over n^2 } ) \sum_{m=1}^\infty { ( m+1) \over m^2 (m^2 - { 1 \over n^2} ) }{ e^{ i m \tau }\over B^m}  \, + c.c. 
\ee
This is the solution to this order. Inserting this into the action we 
can compute the Renyi entropies. We can go to higher orders by solving the conformal welding problem for 
$\theta = \tau + \delta \theta$, as explained in \cite{Mumford}, computing the flux to next order, and solving again
the Schwarzian equation to find the next approximation for $\theta(\tau)$. In this way we can systematically go to any order we want. 

As a check of \nref{thetF}, we can consider the $n\to 1$ limit. In this case all Fourier 
coefficients of  \nref{thetF} go to zero except $m= \pm 1$ so that we get 
\be \la{fis}
\delta \theta = - i { \kappa  \over B} (e^{ i \tau} - e^{ - i \tau } ) 
\ee
In order to compare with the results of the quantum extremal surface calculation we should recall that we have gauge fixed $A$ to be zero. 
Indeed the final solution \nref{fis} looks like an infinitesimal $SL(2,R)$ transformation of the $\theta = \tau$ solution. This is precisely what results from the transformation 
\be
 e^{ i \theta } \sim  e^{ i \tau }(1+ i \delta \theta) \sim  { e^{ i \tau} - A \over 1 - A e^{ i \tau } } \sim e^{ i \tau} ( 1 - A e^{ - i \tau} + A e^{ i \tau } ) ~,~~~~~~~ A \sim { \kappa \over B} \ll 1
 \ee
for small $A$ as in \nref{smallAB}. This shows that the finite-$n$ solution at high temperatures has the right $n\to 1$ limit.


\section{Single interval at zero temperature}
\la{sec:OneZero}

\begin{figure}
\begin{center}
\includegraphics[scale=1.0]{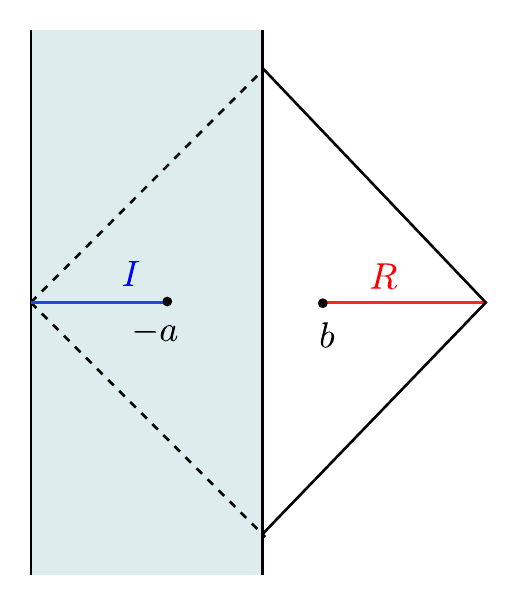}
\end{center}
\caption{\small An information puzzle at zero temperature, with AdS$_2$ on the left and flat space on the right. The naive calculation of matter entropy in region $R$ is infrared-divergent, but this cannot be purified by quantum gravity in AdS$_2$. This is resolved by including the island, $I$.}\label{fig:zerotemp-lorentzian}
\end{figure}

There is a very simple version of the information paradox at zero temperature \cite{Almheiri:2019yqk}. Consider the region $R$ in fig.~\ref{fig:zerotemp-lorentzian}. Ignoring gravity, the von Neumann entropy of the quantum fields on this region is infrared divergent. This is the Hawking-like calculation of the entropy using quantum field theory on a fixed background.

The state of the quantum fields on a full Cauchy slice is pure. However, the AdS$_2$ region is supposed to be a quantum system with $e^{S_0}$ states. This is a contradiction, because it is impossible for the finite states in the AdS$_2$ region to purify the IR-divergent entropy of region $R$. The UV divergence is not relevant to this issue because it is purified by CFT modes very close to the endpoint.

This is resolved by including an island, as in fig.~\ref{fig:zerotemp-lorentzian} \cite{Almheiri:2019yqk}. We will describe briefly how this is reproduced from a replica wormhole. This doesn't require any new calculations because we can take the limit $\beta \to \infty$ in the finite temperature result.  The pictures, however, are slightly different, because the replica geometries degenerate in this limit and the topology changes.

\subsection{Quantum extremal surface}
The metric and dilaton for the zero-temperature solution are 
\be
ds^2_{in} = \frac{4 dy d\by }{(y + \by)^2} \ , \qquad \qquad \phi =- \frac{2\phi_r}{y + \by} \ , ~~~~~~~~y = \sigma + i \tau 
\ee
with $\sigma<0$. As before we glue it to flat space $dy d\bar y $ at $\sigma =0$. The region $R$ and the island $I$ are the intervals
\be
I: \quad y \in (-\infty, -a] , \qquad
R: \quad y \in [b,\infty)
\ee
at $t=0$. 
The generalized entropy, including the island, is
\be
S_{\rm gen}(I \cup R) = \frac{\phi_r}{a} + \frac{c}{6}\log \frac{(a+b)^2}{a} \ .
\ee
Setting $\p_a S_{\rm gen} = 0$ gives the position of the QES,
\be\label{zerotempQES}
a = \frac{1}{2}(k + b + \sqrt{b^2 + 6bk + k^2})  \ , \quad k \equiv \frac{6\phi_r}{c} \ .
\ee

\subsection{Replica wormholes at zero temperature}
\begin{figure}
\begin{center}
\includegraphics[scale=0.5]{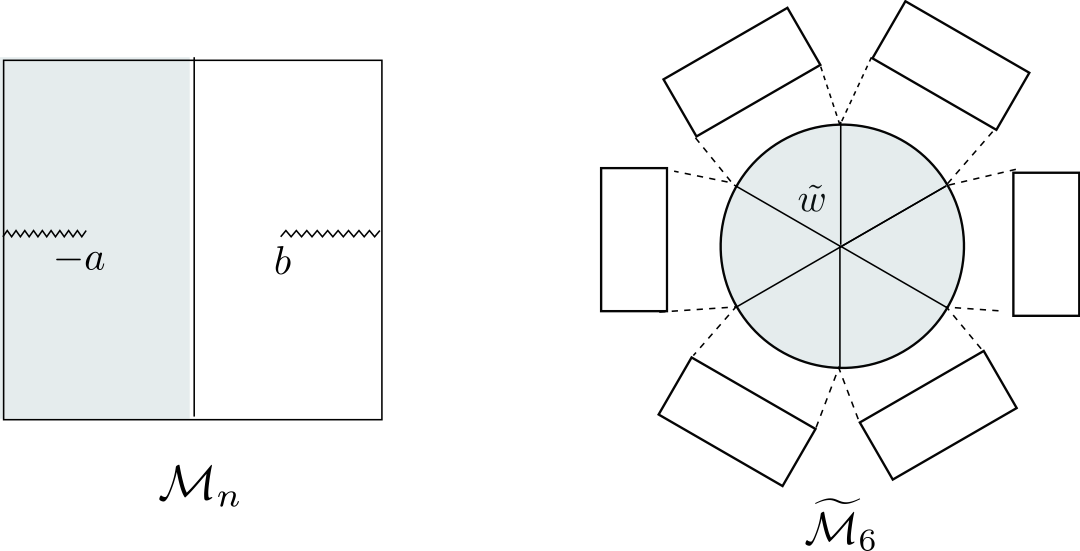}
\end{center}
\caption{\small Replica wormhole at zero temperature. On the right, the disk is glued to $n$ copies of the half-plane, as indicated by the dashed lines. } \label{fig:zerotemp-euclidean}
\end{figure}

The replica partition function $\Tr (\rho_A)^n$ is given by the path integral in fig.~\ref{fig:zerotemp-euclidean}. The boundary condition for the gravity region is $n$ copies of the real line. The Hawking saddle fills in the gravity region with $n$ independent copies of $\mathbb{H}_2$. The replica wormhole, shown in the figure, fills in the gravity region with a single copy of $\mathbb{H}_2$. To see all $n$ sheets of the gravity region, we go to the uniformizing coordinate
\be
\widetilde{w} = \left(\frac{a+y}{a-y } \right)^{1/n} \ .
\ee
This maps the full gravity region to a single hyperbolic disk, $|\widetilde{w}| < 1$. This disk is a wormhole connecting $n$ copies of flat space. The $n^{th}$ copy is glued to the segment with arg $\widetilde w \in [-\frac{\pi}{n}, \frac{\pi}{n}]$.

The equation of motion, and the answer for the position of the QES, is found by taking $\beta \to \infty$ in the results of section \ref{sec:SingleInterval}. This of course agrees with \eqref{zerotempQES}. (It is also possible to solve this problem directly at zero temperature, but we found it easier to treat the welding problem at finite temperature where the gluing is compact. In the end, the welding effects drop out in the determination of the position of the QES, as we saw below \eqref{hilbertp}.)


\section{Two intervals in the eternal black hole}
\label{sec:TwoIntervals}

We now turn to the information paradox in the eternal black hole \cite{Almheiri:2019yqk}, described in the introduction and pictured in fig.~\ref{fig:eternalBH-lorentzian}. In the late-time regime relevant to the information paradox, the generalized entropy, including the island, is simply twice the answer for a single interval. We would like to understand how this is reproduced from wormholes. This is essentially just putting together the general discussion of section \ref{sec:Action} with the single-interval results of section \ref{sec:SingleInterval}, so we will be brief. We will only discuss the saddles near $n=1$; it would be nice to have a more complete understanding of the finite-$n$ wormholes in this setup.

\subsection{Review of the QES}

We set $\beta = 2\pi$. The points in fig.~\ref{fig:eternalBH-lorentzian} have $(\sigma, t)$ coordinates
\be \la{4pts}
P_1 = (-a, t_a) \ , \quad
P_2 = (b, t_b) \ , \quad
P_3 = (-a, -t_a + i \pi) \ , \quad
P_4 = (b, -t_b + i \pi) \ .
\ee
The radiation region is
\be
R = [P_4, \infty_L) \cup [P_2, \infty_R) \ ,
\ee
and the island is
\be
I = [P_3, P_1] \ .
\ee
The CFT state is pure on the full Cauchy slice, so
\be
S_{\rm CFT}(I \cup R) = S_{\rm CFT}( [P_4, P_3] \cup [P_1, P_2] ) \ .
\ee
This entropy is non-universal; it depends on the CFT. In the theory of $c$ free Dirac fermions \cite{Casini:2005rm}, the entanglement entropy of the region
 \be \la{region}
  [x_1, x_2] \cup [x_3, x_4] ~,
 \ee  with metric $ds^2 = \Omega^{-2} dx d\bar x $, is
\be \la{entrexp}
 S_{\rm fermions} 
= \frac{c}{6}\log \left[ \frac{|x_{21} x_{32}  x_{43}  x_{41} |^2}{
  |x_{31}  x_{42} |^2 \Omega_1\Omega_2\Omega_3\Omega_4} \right] \ .
\ee
where we dropped the UV divergences. 
With our kinematics and conformal factors, this gives
\begin{align}\label{bosonee}
S_{\rm fermions}(I \cup R) 
= \frac{c}{3}\log\left[
\frac{ 2\cosh t_a \cosh t_b \left|\cosh(t_a-t_b) - \cosh(a+b)\right| }{ 
\sinh a \cosh (\frac{a+b - t_a - t_b}{2}) \cosh( \frac{a + b + t_a + t_b}{2}) } 
\right]
\end{align}
In a general CFT, the two-interval entanglement entropy is a function of the conformal cross-ratios $(z,\bar{z})$ which agrees with \eqref{bosonee} in the OPE limits $z \to 0$ and $z \to 1$. For concreteness we will do the calculations for the free fermion, but the regime of interest for the information paradox will turn out to be universal.

The generalized entropy, including the island, is
\be\label{SgenFermion}
S_{\rm gen}(I \cup R) = 2S_0 + \frac{2\phi_r}{\tanh a} +   S_{\rm fermions}(I \cup R) \ ,
\ee
Without an island, the entropy is the CFT entropy on the complement of $R$, the interval $[P_4, P_2]$, which is
\be
S_{\rm gen}^{\rm no~island}= S_{\rm fermions}(R) = 
 \frac{c}{3}\log \left(
{2 \cosh t_b } 
\right)
\ee
At $t=0$, 
\be \la{IslandSgen}
S_{\rm gen}^{\rm island}  = 2 S_0 + \frac{2\phi_r}{\tanh a} + \frac{c}{3}\log \left(
\frac{4 \tanh^2 \frac{a+b}{2} }{  \sinh a}
\right) \ .
\ee
The extremality condition $\p_a S_{\rm gen}^{\rm island} = 0$ at $t_a=t_b = 0$ gives
\be \la{eqnFer}
\frac{6\phi_r}{c} \sinh (a+b) = 2 \sinh^2 a - \sinh a \cosh a \sinh(a+b) \ .
\ee
Whether this has a real-valued solution depends on the parameters $b$ and $\phi_r/c$. For example, if $b=0$, then it has a real solution minimizing $S_{\rm gen}^{\rm island}$ when $\phi_r/c$ is small, but not otherwise. 

At late times, the extremality condition $\p_a S_{\rm gen}^{\rm island} = 0$ always has a real solution. The true entropy, according to the QES prescription, is 
\be
S(R) =
\min\left\{ S_{\rm gen}^{\rm no ~island} \ ,   S_{\rm gen}^{\rm island}   \right\} \ .
\ee
The island always exists and dominates the entropy at late times, because the non-island entropy grows linearly with $t$, see   fig.~\ref{Page}. This solution is in the OPE limit where we can approximate the entanglement entropy by twice the single-interval answer,
\be
S_{\rm matter}(I \cup R) \approx 2 S_{\rm matter}^{\rm   } ([P_1, P_2]) =
 \frac{c}{3}\log \left(
\frac{2|\cosh(a+b) - \cosh(t_a-t_b)|}{
\sinh a} 
\right) \ .
\ee
and the QES condition sets $t_a = t_b$. 

\subsection{Replica wormholes}

We would like to discuss some aspects of the wormhole solutions that lead to the island prescription. 

For general $n$ these are wormholes which have the topology shown in figures \ref{fig:2wormhole}(b), \ref{fig:manyreplicas}. 
Already from these figures we can derive the  $S_0$-dependent contribution \nref{OrigAct} 
since it involves only the topology of the manifold. The replica wormhole that involves nontrivial connections, 
see figure   
\nref{fig:manyreplicas}, has the topology of a sphere with $n$ holes. This gives a contribution going like 
$Z_n \propto e^{ S_0( 2-n)}$ and a contribution of $2S_0 = (1 - n\partial_n) \log Z_n   |_{n=1}$ \,  for the von Neumann entropy. This is good, since the island contribution 
indeed had such a term \nref{IslandSgen}.

It is useful to assume replica symmetry and view the Riemann surface as arising from a single disk with 
$n$ copies of the matter theory and with pairs of twist operators that connect all these $n$ copies in a cyclic fashion, see figure \ref{CoverThree}. In order to find the full answer, we need to solve the equations \nref{EOMFin} \nref{PartSing}. The important point is that, at this stage, we have that $n$ appears purely as a parameter and we can analytically continue the equations in $n$. 
  We have not managed to solve the equations for finite $n$. 
But let us discuss some properties we expect. 
In the limit of large $c \beta/\phi_r$, it is likely that solutions exist in Euclidean signature.\footnote{For low values of $c\beta/\phi_r$ we have already seen, in \nref{eqnFer}, that near $n\sim 1$ the solutions can be complex.} 
We can put points 
$P_2$ and $P_4$ at  $v=\pm B e^{ \pm i \varphi }$.   Once this solution is found, we can analytically continue $\varphi \to - i t$ to generate the Lorentzian solution. 
That Lorentzian solution at late times $t$ is expected to exist even for low values of $c \beta/\phi_r$.
In principle, it should be possible, and probably easier,  to analyze directly the late-times Lorentzian equation. 
In fact, 
 we expect that there should be a way to relate the single interval solution to the two interval solution in this regime. The intuitive reason is that at late times the distance between the two horizons is increasing and so the distance between the two cosmic branes is increasing.  We have an external source cosmic brane outside the gravitational region, at the tip of region $R$. 
 The cosmic brane has some tension, as well as a twist operator on it. For the Hawking saddle, the one without the replica wormholes, the twist operators, and the topological line operators\footnote{These topological line operators exchange the $n$ copies in a cyclic way. They are represented by red lines in figure \ref{CoverThree}(b).} that connect them, generate a contribution that grows linearly in time, due to the behavior of 
 Renyi entropies for the matter quantum field theory, as well as the fact that the wormhole length grows with time. 
   At late times the topological line operator can break by pair producing cosmic branes, with their twist operators. The cost of creating a pair of cosmic branes is finite in the gravitational region, because the dilaton is finite. This cost would be infinite in the non-gravitational region. 
  But once the external cosmic brane is screened by the cosmic brane that appeared in the gravity region we expect to have two approximately  independent single interval problems. The reason is that the distance between the left and right sides is growing with time.
  This is somewhat analogous to two point charges that generate a two dimensional electric field. As one separates the charges it might be convenient to create a pair of charges that screens the electric field. For this it is important that the charges one creates have finite mass.

 In the $n\to 1$ limit we can analyze the solution and we get the generalized entropy. This is not too surprising since the arguments in \cite{Dong:2017xht} say that this should always work. Here the non-trivial input is the ansatz for the configuration of intervals which follows from the structure of the Riemann surfaces. As discussed in section \ref{sec:CosmicStrings}, the effective action reduces to the action of certain cosmic  branes which are manifestly very light in the $n\to 1$ limit. So in this case, the argument of the previous paragraph can be explicitly checked and one indeed obtains that we get the sum over the two single interval problems \cite{Almheiri:2019yqk}.

\subsection{Purity of the total state}

One can take the perspective that our model is defined via a quantum theory living on the flat space region including its boundary endpoints. The global pure state we consider should be a pure state of this region, and a natural question is whether this is captured in the gravity description. Replica wormholes do indeed capture this feature.

\begin{figure}
\begin{center}
\includegraphics[scale=0.3]{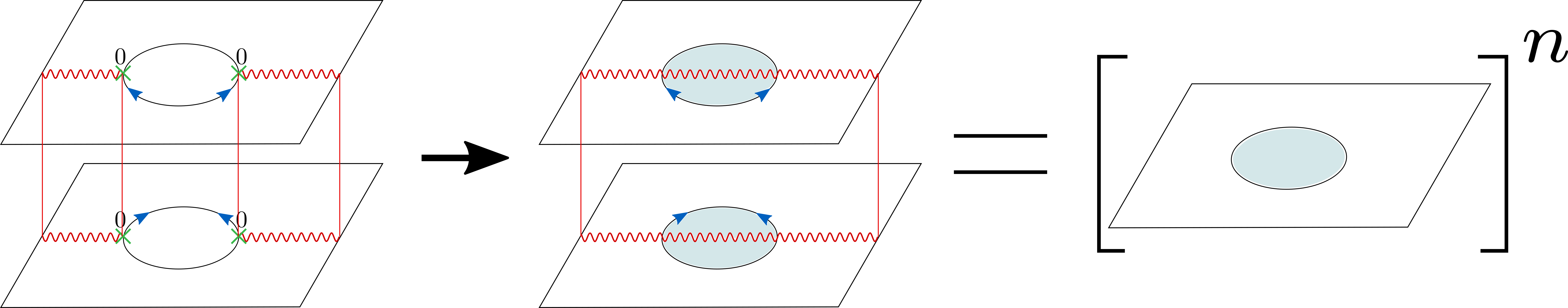}
\end{center}
\caption{\small The computation of the entropy of the entire flat space regions including the boundary points. The dominant gravitational saddle connects consecutive sheets. This factorizes into $n$ separate sheets and produces a vanishing entropy consistent with the purity of the flat space region union the endpoints. The blue arrows indicate how the unit circle is identified across the cut.
\label{FullIntervalSheets}}
\end{figure}

The computation of the entropy of this region is given by evaluating the path integral on the manifold shown in figure \ref{FullIntervalSheets}. The branch cuts split the entire flat space region including its boundaries, identifying one half of one sheet with the other half of the next sheet. The most obvious gravitational saddle is the one that connects these consecutive sheets and thereby naturally extending the branch cut through the entire gravity region. A simple rearranging of these sheets shows that this contribution to the Renyi entropy factorizes. This disconnected saddle satisfies $Z_n = Z_1^n$, and evaluating the on shell action on this configuration will give vanishing entropy since
\begin{align}
\Tr \rho^n = {Z_n \over Z_1^n} = 1\,.
\end{align}
This saddle clearly dominates over all other configurations. 

Since the different sheets are not coupled at all in the flat space region, it's plausible that this disconnected saddle is the only saddle that exists. Other off-shell contributions can indeed exist, but we speculate they should give a vanishing contribution in a model with a definite Hamiltonian with no averaging.


\section{Comments on reconstructing the interior}
\label{sec:Dictionary}

\begin{figure}
\begin{center}
\includegraphics[scale=0.6]{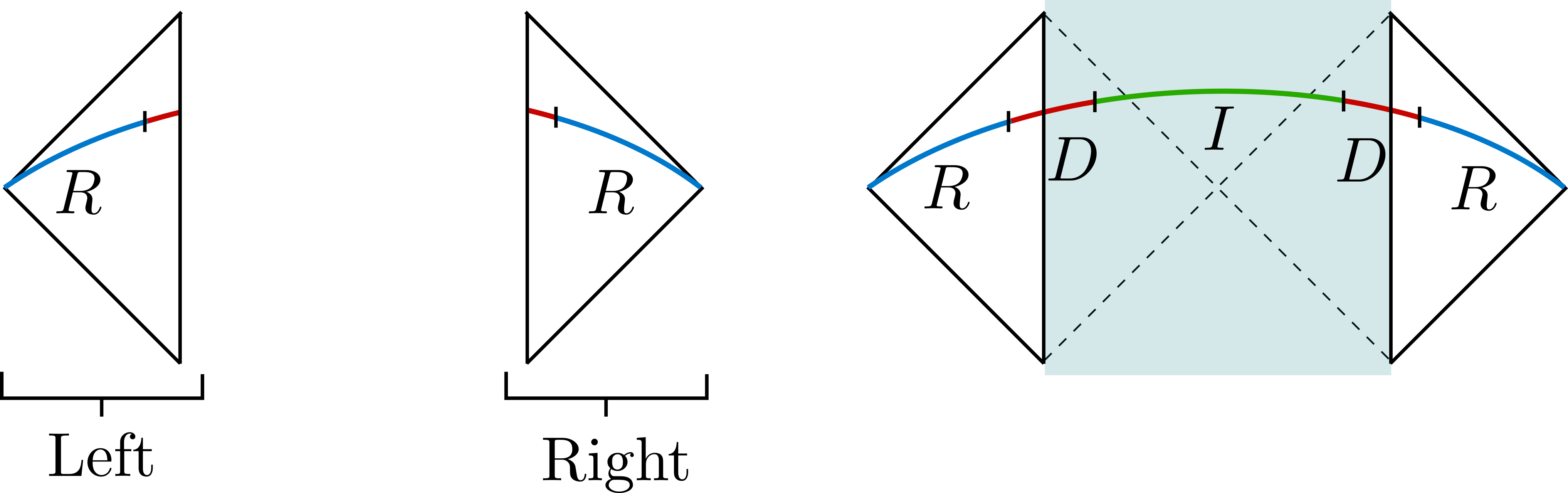}
(a) ~~~~~~~~~~~~~~~~~~~~~~~~~~~~~~~~~~~~~~~~~~~~~~~~~~~~~~~ ~~~~~(b)
\end{center}
\caption{\small (a) The full Hilbert space is the product Hilbert space of the entire left and right flat space regions including the boundary points. The region $R$ we are interested in is a union of two subregions in the two flat space regions. (b) The effective state used in the island prescription is the semi-classical state defined on the Cauchy slice of the full system. $R$ is the same region in the flat space region whose exact entropy we are computing, $I$ is the island, and $D$ is the complement of the two.
\label{JLMSlatetime}}
\end{figure}

The island contribution to the entropy of a flat space region $R$ indicates there is a dictionary between the island $I$ and $R$ in the sense of entanglement wedge reconstruction in AdS/CFT. We could discuss this in general but for concreteness consider the two interval case discussed in the previous section. Let's take the state at late times such that the entropy of $R$ has plateaued and its entropy receives a contribution from the island as shown in figure \ref{JLMSlatetime}.

The first step to establishing a dictionary is to define a subspace of states which have the same ``entanglement wedge'' or island. This defines what we will call the code subspace  ${\cal H}_{code}$, which we imagine can be prepared via the Euclidean path integral with possible operator insertions. By having the same island we mean that the leading saddle points in the Renyi computations are only modified perturbatively. This naturally puts restrictions on the size of the allowed code subspace for which the statements of this section hold, see for example \cite{Hayden:2018khn, hayden2017approximate}.

We assume that the full Hilbert space of our model is that of the two flat space regions including their boundary, which we write as $ {\cal H}_{\rm Left} \otimes {\cal H}_{\rm Right}$. The region $R$ that we are considering is a tensor factor of this Hilbert space, where we can write
\begin{align}
 {\cal H}_{\rm Left} \otimes {\cal H}_{\rm Right} = {\cal H}_R \otimes {\cal H}_{\bar R}
\end{align}
where $\bar{R}$ is the complement of the region $R$ in the flat space region including the boundary points. 

The code subspace ${\cal H}_{code}$ is a subspace of $ {\cal H}_{\rm Left} \otimes {\cal H}_{\rm Right}$. However, the code subspace also has a simpler description in terms of the combined description of gravity plus the flat space region as that of effective field theory on a Cauchy slice of the full spacetime. This is the description where the state is prepared using the semi-classical saddle via the Euclidean black hole solution. 
The code subspace should be thought of as isomorphic to this. Therefore, the code subspace admits the decomposition\footnote{This should be understood as approximate up to usual issues of the non-factorizability of continuum QFT.}
\begin{align}
{\cal H}_{code} \cong {\cal H}_{R} \otimes {\cal H}_{D} \otimes {\cal H}_{I} 
\end{align}
where the region $D$ is the complement of $R \cup I$ on the Cauchy slice. The decomposition is shown in figure \ref{JLMSlatetime}. It is within this effective description that for any state in the code subspace $|i \rangle \in {\cal H}_{code}$, we have
\begin{align}
S({\bold \rho}_{R}^{i}) = S(\tilde{\rho}_{RI}^{i}) + {\mathrm{Area}[\partial I] \over 4 G_N}
\end{align}
where $\rho^i_{R}$ is what you get by tracing out $\bar{R}$ in the full quantum description ${\cal H}_{\rm Left} \otimes {\cal H}_{\rm Right}$, and $\tilde{\rho}_{RI}^i$ is the density matrix obtained by tracing out the complement of $RI$ in the semi-classical description consisting of quantum fields on a classical geometry.

The validity of the island formula (for a fixed island) within the code subspace implies the equivalence of the relative entropy in the exact state and the semi-classical state:
\begin{align} \la{RelEnt}
S_\mathrm{Rel}(\rho_{R} | \sigma_{R}) = S_\mathrm{Rel}(\tilde{\rho}_{ R I} | \tilde{\sigma}_{ R I})
\end{align}
A similar observation in the context of AdS/CFT \cite{Jafferis:2015del} was key in proving entanglement wedge reconstruction \cite{Dong:2016eik} using the quantum error correction interpretation of the duality\cite{Almheiri:2014lwa}. The same line of argument can be applied here to establish the dictionary. In particular, one can show that for any operator ${\cal O}_I$ (and its Hermitian conjugate) acting within the ${\cal H}_{code}$ and supported on the island one can find an operator supported on $R$ such that:
\begin{align}
{\cal O}_I | i \rangle  &= {\cal O}_{R} | i \rangle \\
{\cal O}_I^\dagger | i \rangle  &= {\cal O}_{R}^\dagger | i \rangle
\end{align}
The operator $ {\cal O}_{R}$ is given by a complicated operator on ${R}$ involving the matrix elements of ${\cal O}_I$ within the code subspace.

In summary, we are using the fine grained entropy formula to understand how the interior is encoded in the full Hilbert space. The relative entropy equality \eqref{RelEnt} tells us that distinguishable states in the interior (the island) are also distinguishable in the radiation, within the full exact quantum description.

\section{Discussion}
\la{sec:Conclusions} 


In this paper, we have exhibited non-perturbative effects that dramatically reduce the late time von Neumann entropy of quantum fields outside a black hole.

The computation of the Renyi entropies corresponds to the expectation value of a swap or cyclic permutation operator in $n$ copies of the theory. Systems with very high entropy have very small, exponentially small,  expectation values for this observable. This means that non-perturbative effects can compete with the naive answers. In particular, the Hawking-like computation of the Renyi entropies of radiation corresponds to a computation on the leading gravitational background. A growing entropy corresponds to an exponentially decreasing expectation value for the cyclic permutation operator. It decreases exponentially as time progresses. 
 For this reason, we need to pay attention to other geometries, with other topologies. These other topologies give exponentially small effects, but they do not continue decreasing with time for long times. 
 Said in this way, the effects are vaguely similar to the ones discussed for corrections of other exponentially small effects 
 \cite{Maldacena:2001kr,Saad:2018bqo,Saad:2019lba,Saad:2019pqd}. 
 Though the Renyi entropies are small,  the von Neumann entropy is large and  the new series of saddles gives rise to a constant von Neumann  
 entropy at late times.  More precisely, we can think of the computation of the Renyi entropies in the two interval case as an insertion of a pair of external cosmic branes in the non-gravitational region. As time progresses these are separated further and further through the wormhole. Eventually the dominant contribution is one where a pair of cosmic branes is created in the gravitational region that ``screen'' the external ones, giving an entropy which is the same as that of two copies of the single interval entropy. 
 
   These other topologies are present as subleading saddles also at short times (perhaps as complex saddles) where we can analyze them using Euclidean methods and then analytically continue. We have only done this analytic continuation for the von Neumann entropies, not the Renyi entropies. It would be interesting to do it more explicitly for the Renyi entropies.
  
  There have been discussions on whether small corrections to the density matrix, of order $e^{ -S_{BH}}$,
 could or could not restore unitarity. These results suggest that they interfere constructively to give rise 
 to the right expression for the entropy. 
  
  This is evidence that including  nonperturbative gravitational effects can indeed lead to results compatible with unitarity.   However, we emphasize that this is not a full  microscopic resolution of the information paradox. We have not given a gravitational description for   the 
   $S$-matrix describing how infalling matter escapes into the radiation. 
   In this sense, these  results are on a footing similar to the Bekenstein-Hawking calculation of the entropy, which uses a Euclidean path integral to compute the right answer but does not give an explicit  Hilbert space picture for what it is counting. 
   In contrast, the Strominger-Vafa computation of the entropy \cite{Strominger:1996sh} gives us an explicit Hilbert space, but not a detailed description of the   microstates in the gravity variables.
    Something similar can be said of the CFT description in AdS/CFT. Hopefully these results will be useful for providing a more explicit map. 
   
   It is amusing to note that wormholes were initially thought to destroy information 
   \cite{Hawking:1987mz,Lavrelashvili:1987jg,Giddings:1987cg}. 
   But more recently the 
   work of \cite{Saad:2019lba,Saad:2019pqd},  as well as the present discussion, and \cite{Penington:2019kki}, suggests that the opposite is true. Wormholes are important for producing results that are compatible with unitarity. 
   For earlier work in this direction see also    \cite{Coleman:1988cy,Giddings:1988cx,Polchinski:1994zs}.

We assumed that $c\gg 1$ as a blanket justification for analyzing the equations classically. However, even for small $c \sim 1$, the basic picture for the Page curve can be justified. The basic point is simple. First consider the single interval computation. In that case for $c\sim 1$ we see that the correction to the black hole solution is very small, for all the Renyi entropies. In other words, we find that $A$ is small, and we can probably not distinguish such a small value of $A$ from zero but that does not matter, the geometries and the entropies are basically those of a black hole. Now when we go to two intervals, and we consider the late time situation, then all that really matters is that we can do an OPE-like expansion of the twist operator insertions. The important observation is that the twist operator insertions in the interior of the black hole are very far from each other. This is the fact that the wormhole is getting longer \cite{Hartman:2013qma,Susskind:2014moa}. Then the solution becomes similar to two non-interacting copies of the single interval solution.
The fact that $c$ is small only implies that we will have to wait longer for the island solution to dominate.  We just have to wait a time of order the entropy, $t \propto  {\beta } (S-S_0)/c$ for it to dominate.  

In \cite{Saad:2019lba}, it was argued that pure JT gravity should be interpreted in terms of an average over Hamiltonians. In addition, higher genus corrections were precisely matched. 
This has raised the question of whether the corrections we are discussing in this paper crucially involve an average over Hamiltonians, or whether they would also apply to a system which has a definite Hamiltonian. 
Though JT gravity plus a CFT probably does not define a complete quantum gravity theory, it seems likely that well defined theories could be approximated by   JT gravity plus a CFT. For example, we could imagine an AdS/CFT example that involves an extremal black hole such that it also has a CFT on its geometry. All we need is this low energy description, the theory might have lots of other massive fields which will not drastically participate in the discussion. They might lead to additional saddles, but it seems that they will not correct the saddles we have been discussing. And we have the seen that the saddles we discussed already give an answer consistent with unitarity, at least for the entropy.  In contrast with \cite{Saad:2019lba}, we are not doing the full path integral, we are simply using a saddle point approximation, so the JT gravity plus CFT only needs to be valid around these saddles.

  As we mentioned in the introduction, the setup in this paper can be 
  viewed as an approximation to some magnetically charged near extremal four dimensional black holes \cite{Maldacena:2018gjk}. 
  But one could analyze more general asymptotically flat black holes and wonder how to define either exactly or approximately the various entropies involved. In particular, to have a sharp definition of the entropy of radiation it seems important to go to null infinity. 
    
  Another interesting question is whether we can give a Lorentzian interpretation to the modification of the density matrix implied by the existence of replica wormholes.  
    
 It has been pointed out that a black hole as seen from outside looks like a system obeying the laws of hydrodynamics. For this reason, it is sometimes thought that gravity is just an approximation that intrinsically loses information. Here we see that if we include the black hole interior, and we do a more complete gravity computation, we can get results compatible with unitarity. The fact that gravity is more than dissipative hydrodynamics is already contained in the Ryu-Takayanagi formula for the fine grained entropy, which shows that the geometry of the interior can discriminate between pure and mixed states for a black hole.

\vspace{1cm}
\textbf{Acknowledgments} We are grateful to  
Tarek Anous, Raphael Bousso, Kanato Goto, Daniel Harlow, Luca Iliesiu, Alexei Kitaev, Alexandru Lupsasca, Raghu Mahajan, Alexei Milekhin, Shiraz Minwalla,  Geoff Penington, Steve Shenker, Julian Sonner, Douglas Stanford, Andrew Strominger, Sandip Trivedi and Zhenbin Yang 
for helpful discussions of this and related work. 
AA and TH thank the organizers of the workshop \textit{Quantum Information In Quantum Gravity} at UC Davis, August 2019, and AA, TH, and JM thank the organizers of the workshop \textit{Quantum Gravity in the Lab} at Google X, November 2019. A.A. is supported by funds from the Ministry of Presidential Affairs, UAE.
The work of ES is supported by the Simons Foundation as part of the Simons Collaboration on the Nonperturbative Bootstrap. The work of TH and AT is supported by DOE grant DE-SC0020397.
J.M. is supported in part by U.S. Department of Energy grant DE-SC0009988 and by the Simons Foundation grant 385600.

\appendix


\section{Derivation of the gravitational action} 
\la{GravAct}

In this appendix we derive the action that leads to the equation of motion 
\nref{EOMFin}. 

We start with the expansion of the metric near the boundary  \nref{Metdr}
  \nref{DelRhoe}
\be
ds^2 = { 4 dw d\bar w \over (1 - |w|^2)^2}  \left(1 - { 2 \over 3} (1-|w|)^2 U(\theta) + \cdots  \right).
\ee
We now write in terms of the variables $w = e^{-\gamma} e^{ i\theta }$ and expand it in powers of $\gamma$ as 
\be
ds^2 = { d \theta^2 \over \gamma^2 } +  { d \gamma^2 \over \gamma^2 } - { 2 \over 3 } d\theta^2 U(\theta) \,.
\ee
We now equate this to $ds = { d\tau \over \epsilon}$, we set $\theta = \theta(\tau)$  and solve 
for $\gamma$ in a power series 
\be
\gamma = \epsilon  \theta' \left[1  + \epsilon^2 \left( \half {  {\theta''}^2 \over { \theta'}^2 } - { 1 \over 3}
 U(\theta) {\theta'}^2 \right) + \cdots \right].
\ee
We can now compute the tangent vector to the curve $t^\mu$ and the normal vector $n^\mu$ and compute the
extrinsic curvature from 
\be
K = t^\mu t^\nu \nabla_\mu n_\nu = 1 + \epsilon^2 \left[ \{ \theta , \tau \} +  \left( \half + U(\theta)\right) {\theta'}^2 \right].
\ee
Up to the purely topological term, the gravitational action \nref{newgra}
 reduces to the extrinsic curvature term 
 \be
 - I_{\rm grav} = { 1 \over 4 \pi }  { \phi_r \over \epsilon } \int { d \tau \over \epsilon } 2 K =  { 2 \phi_r \over 4\pi \epsilon^2} \int d\tau + {\phi_r \over 2 \pi }  \int d\tau \left[ \{ \theta(\tau) , \tau\} +  \left( \half + U(\theta)\right) {\theta'}^2 \right] + o(\epsilon)\,.
 \ee
 The first term is a purely local divergence that can be viewed as the correction to the vacuum energy. 
 We should also remark that we can always choose a coordinate $x$ where the metric locally looks like the standard Poincare coordinates. In those coordinates the action is simply $\{x,\tau\}$. However, we will have a nontrivial identification for $x$ as we move from $\tau \to \tau + 2\pi $. Here we simplified the boundary condition, it is just $\theta = \theta + 2 \pi$, but we complicated a bit the action. Notice that
 we can think of $U(\theta)$ as a stress tensor, the change of coordinates is basically the same that we use to 
 transform this stress tensor to zero. In other words, $x(\theta)$ is a function which obeys $\{x,\theta \} = \half + U(\theta)$. 
 
The conserved energy of the system is given by 
 \be
  E = {\phi_r \over 2 \pi }    \left[ \{ \theta(\tau) , \tau\} +  
  \left( \half + U(\theta)\right) {\theta'}^2 \right] .
  \ee

  We now compute $U(\theta)$ for the case when  we put a conical defect at point $A$ in the $w$ plane. We have
  the metric \nref{metrictil} and the change of coordinates \nref{wtildew} which imply that 
  \bea
  ds^2 &=& \left| { d \tilde w \over d w } \right|^2 { 4 |dw|^2 \over (1 - |\tilde w|^2 )^2 }  
  \cr 
   &=& { 4 |dw|^2 \over (1 - |w|^2)^2 } \left[ 1 - { 2 \over 3} (1 - |w|)^2 U(\theta)  + \cdots \right] ~,~~~{\rm as} ~~|w|\to 1 
  \eea
  with 
  \be
   U(\theta) = -\half \left( 1 - { 1 \over n^2} \right)  \frac{(1-A^2)^2   }{ (e^{i\theta}  - A)^2(e^{-i \theta}  - A  )^2}   \,,
   \ee
   which leads to the same action as \nref{Rdefi}
  
   We now would like to derive the equations of motion for this action. In particular, we would like to see that
 as $\theta \to \theta + \delta \theta$ we get the right equations of motion. The change in gravitational action is 
 simple, we just have 
 \be \la{VarSch}
  - \delta I_{\rm grav} = - { \phi_r \over 2 \pi } \int d\tau  { \left[  \{ \theta(\tau) , \tau\} +  ( \half + U(\theta)) {\theta'}^2 \right]' \over \theta' }\, \delta \theta \,.
  \ee

  Now, let us do the variation of the CFT part. 
  Imagine that we choose locally complex coordinates so that 
   \be 
   \log w = s + i \theta  
   \ee
   We also have the outside coordinates $y = \sigma + i \tau$ and we can locally think of the relation between the two in terms of $\log w = i \theta(-iy)$. 
   Now imagine that we do a small change $\theta(\tau) \to \theta + \delta \theta$ with $\delta \theta$ with compact 
   support. This would change the relation between the two sides. However, let us imagine we instead keep the relation fixed, set by $\theta(\tau)$ and we redefine the outside coordinate by an infinitesimal  reparametrization, 
   $ \tilde y = y + \zeta^y $ in such a way that the relation between the new variables is the same as the old one
   \be \la{DefTvf}
   \log w = i \theta (-i \tilde y) = i \theta(-i y) + i \delta \theta(-iy) = i \theta(-iy) + \theta'(-iy) \zeta^y  ~~~~~~\to ~~~~ 
   \zeta^y = i { \delta \theta \over \theta' } 
   \ee
   and we have the complex conjugate expression for $\zeta^{\bar y}$. 
   We can then extend this reparametrization in a non-holomorphic way in the region outside, defining 
   \be
    \tilde \zeta^y = i { \delta \theta(-i y) \over \theta'(-iy ) } h(\sigma) ~,~~~~~~~~ \tilde \zeta^{\bar y} =- i 
   { \delta \theta(i \bar y ) \over \theta'(i \bar y ) } h(\sigma)
   \ee
   where $h(\sigma)$ is one for $\sigma =0$ and quickly goes to zero at $\sigma $ increases. An example is 
   $h(\sigma ) = \theta(\sigma_0 -\sigma )$ for a small $\sigma_0$. 
   This change of coordinates is equivalent to a change in metric 
   \be
   ds^2 = dy d\bar y = d\tilde y d\bar {\tilde  y} - 2 \partial_{\alpha } \zeta^{\beta} d{\tilde y}^\alpha d{\tilde y}^\beta ~,~~~~~
   \delta g_{\alpha \beta } = - 2 \partial_{(\alpha } \zeta_{\beta)} 
   \ee
  This differs from the original metric by some terms that are localized near the point where we are doing the variation.   The relation between $\log w$ and
   the $\tilde y$ variable was the same as it was before we did the variation, due
   to our choice of $\tilde y$ variable in \nref{DefTvf}. Furthermore, far from the
   region where we are doing the variation, both variables coincide. 
  Thus, the only thing we are doing is locally changing the metric of the outside region. 
Using the definition of the stress tensor, $T_{\alpha \beta} = - { 2 \over  \sqrt{g} } { \delta  \over \delta g^{\alpha \beta } } \log Z $, we get  
  \bea  \la{InteRes}
  \delta \log \hat Z_M &=& 
  -
  \half 
  \int d\varphi d \sigma ( T_{yy }   \delta g^{yy} +      T_{\bar y \bar y }  \delta g^{\bar y \bar y } )
 \cr 
  &=&
 - 2
  \int d\varphi d \sigma ( T_{yy }  \partial_{\bar y}  \zeta^{y}  + T_{\bar y \bar y }  \partial_{  y }  \zeta^{\bar y} )\,,
  \eea
  where we used that the background metric is flat and that the trace of the stress tensor is zero. 
     We now use evaluate the derivatives 
  \be
  \partial_{\bar y} \zeta^{y } = { i \over 2 } { \delta \theta(-i y ) \over \theta'(-i y) } h'(\sigma)   ~,~~~~~~~~~~  \partial_{  y} \zeta^{\bar y}= -{ i \over 2 }  { \delta \theta (i\bar y ) \over \theta'(i\bar y) } h'(\sigma) ~,~~~~~~~~~~~~~ h' = - \delta(\sigma -\sigma^0)\,.
\ee
Here we used that the arguments of $\delta \theta$ and $\theta'$ are holomorphic or antiholomorphic, so the derivative receives only a contribution from $h$, which is just a delta function. Inserting this into \nref{InteRes}, integrating over $\sigma$, and taking $\sigma^0\to 0$, we get 
\be
\delta \log \hat Z_M =    i  \int d\tau ( T_{yy } -   T_{\bar y\bar y } ) { \delta \theta \over \theta' }   \,.
\ee
Using \nref{VarSch} we get the appropriate equation \nref{EOMFin} after cancelling the $1/\theta'$ factor
from both sides.


\section{Linearized solution to the welding problem}
\label{app:welding}

Let us start with a discussion of the symmetries of the welding problem   \nref{WeldEqns}. 
First we can imagine doing $SL(2,C)$ transformations of the $z$ plane. These move around the point at infinity, and we would need to   allow a pole  in the functions $F$ or $G$. If we fix that $F(\infty ) = \infty$, then we can then impose
that the functions are holomorphic everywhere, with no poles,  and this group is reduced to just translations, scalings and rotations of the plane $z$. 
None of these transformations change the data for the welding problem which is $\theta(\tau)$. 
In addition, we have two $SL(2,R)$ transformations, one acting on $w$ and one acting on $v$, both preserving the circles 
$|w|=1$ and $|v|=1$. These change the data 
of the welding problem by an $SL(2,R)$ transformation of $e^{ i \theta}$ or $e^{i \tau}$ respectively. They map a solution of a 
welding problem with $\theta(\tau)$ to a solution of a different welding problem given by the transformed function. 
In our combined gravity plus CFT problem, we are integrating over $\theta(\tau)$, so we can look for symmetries that change $\theta(\tau)$.  It turns out that the $SL(2,R)_v$ that acts on the $v$ plane is {\it not} a symmetry. It changes the 
Schwarzian action, for example. On the other hand, the $SL(2,R)_w$ is actually a    gauge symmetry, when we also act with the
$SL(2,R)$ transformation on the possible locations, $w_i$, of the conical singularities.

Consider a plane with coordinate $w$ inside the unit disk, and $v$ outside, as in fig.~\ref{WandZplanes}. The plane is glued along the unit circle with a gluing function $\theta(\tau)$, where $w = e^{i\theta}$ and $v = e^{i\tau}$. The solution to the welding problem is a pair of functions
\begin{align}
z &= G(w) \qquad (\mbox{inside})\\
z &= F(v) \qquad (\mbox{outside})
\end{align}
where $G$ is holomorphic inside the disk, and $F$ is holomorphic outside the disk.  In this appendix we will solve for $F,G$ perturbatively, assuming the gluing is close to the identity, $\theta(\tau) = \tau + \delta \theta(\tau)$. Here we are considering $\delta \theta(\tau)$ to be a fixed input to the problem of finding $F$ and $G$.

Expand in Fourier modes,
\begin{align}
\theta(\tau) = \tau  + \sum_{m=-\infty}^{\infty} c_m e^{i m \tau} \ , \quad
G(w) = w + \sum_{\ell=0}^{\infty} g_\ell w^{\ell}  \ , \quad
F(v) =  v  + \sum_{\ell=-\infty}^2 f_\ell v^{\ell} \ .
\end{align}
Here $c_m, d_1^{\ell}$, and $d_2^{\ell}$ are considered small. There is an SL(2) ambiguity in the zeroth order solution, which we have gauge-fixed to set these maps to the identity. (Note that this is different from the choice in the main text around eqn \eqref{Fleading}.) The matching condition on the unit circle is
\be
G(e^{i \theta(\tau)}) = F(e^{i\theta}) \ .
\ee
At the linearized level, this sets
\begin{align}
 f_{\ell+1} &= i c_\ell\qquad (\ell \leq -2)\\
g_{\ell+1} &=  -i c_\ell \qquad (\ell \geq 2) \end{align}
and
\begin{align}
ic_{-1} =  f_2 -g_2 
 \ , \quad
ic_0 = f_1 -g_1
\ , \quad 
ic_1 = 
f_2 -g_2 \ .
\end{align}
There an ambiguity by a small  $SL(2,C)$ action on the $z$ plane.   
We can fix it by setting $G(0)=0$, $F(v) = v + $constant, as $v\to \infty$. This amounts to three complex conditions 
that set 
\be
g_0=f_{2} = f_1 =0 
\ee
This now implies that we get a unique solution for the remaining coefficients in terms of the $c_m$ 
%
\begin{align}
f_l  = i c_{\ell-1} ,   ~~~~{\rm for}~~\ell \leq 0 ~;  \qquad ~~~~~~~~
g_\ell = - i c_{\ell-1} ~,~~~{\rm for}~~~ \ell > 0  \ .
\end{align}
From here we can calculate
\be
v^2 \{ F, v \} =  \sum_{\ell=-\infty}^{-2} \ell(\ell^2-1)i c_\ell v^{\ell } \ .
\ee
Comparing to $\{w, \tau\} = \{ e^{i\theta}, \tau \}$ gives the relation used in the main text,
\be
e^{2i\tau}\{F, v \} = - \delta\{w, \tau\}_- = - ( \delta \theta''' + \delta \theta' )_- \ .
\ee


\section{The equation of motion in Lorentzian signature}
\label{app:lorentzian}

The Hilbert transform appearing in the equation of motion \eqref{hilbertp} has a nice interpretation in Lorentzian signature. It is responsible for the dissipation of energy into the thermal bath outside. This makes contact with the Schwarzian equation for black hole evaporation studied in \cite{Engelsoy:2016xyb,Almheiri:2019psf}.

In this appendix we set $n=1$, but allow for CFT operators inserted in the non-gravitational region. The perturbative Schwarzian equation in Euclidean signature is
\be
\p_\tau S + i \kappa \Hilbert \cdot S = i \kappa {\cal F}
\ee
where $S = \delta \{e^{i\theta}, \tau\}$ and
\be
{\cal F} =  T_{yy}(i\tau) - T_{yy}(-i\tau)  \ .
\ee
We separate this into positive and negative frequencies on the Euclidean $\tau$-circle,
\begin{align}
\p_\tau S_+ - i  \kappa S_+ &= i \kappa {\cal F}_+ \\
\p_\tau S_- + i  \kappa S_- &=  i\kappa {\cal F}_- \ .
\end{align}
Here the `$+$' terms include only the non-negative powers of $e^y$, and the `$-$' terms have the negative powers.
Now continuing to Lorentzian signature with $\tau = it$, this becomes
\begin{align}\label{lorentzianshock}
\p_t S_\pm \pm \kappa S_\pm = -\kappa {\cal F}_\pm
\end{align}
This is the Lorentzian equation of motion. As an example, consider a state with two scalar operators ${\cal O}(y_1) {\cal O}(y_2)$ inserted at 
\be
y_1 = L + i \delta , \qquad y_2  = \by_1 =  L - i \delta \ ,
\ee
with $0 < \delta \ll L $. This creates a shockwave that falls into the AdS region at time $t \approx L$. The state is time-symmetric, so there is also a shockwave exiting the AdS region at $t \approx -L$.
The stress tensor is 
\be
T_{yy}(y) = -  \frac{h_O}{2\pi}  \frac{v^2(v_1 -v_2)^2}{(v-v_1)^2(v-v_2)^2} \ ,
\ee
with $v = e^{y}$. The projections onto positive and negative Euclidean frequencies are
\be
{\cal F}_+ = - \frac{h_O}{2\pi} \frac{v^2(v_1 - v_2)^2}{(v-v_1)^2(v-v_2)^2}  \ , \qquad
{\cal F}_-  = \frac{h_O}{2\pi} \frac{v^2(v_1-v_2)^2}{(1-v_1 v)^2(1-v_1 v)^2} \ .
\ee
In Lorentzian signature this becomes
\begin{align}
{\cal F}_+ &=  \frac{h_O\sin^2\delta }{2\pi (\cos\delta  - \cosh(L+t))^2} \\
{\cal F}_- &= -\frac{h_O \sin^2\delta }{ 2\pi (\cos\delta - \cosh(L-t))^2 } 
\end{align}
As $\delta \to 0$, these vanishes away from the singularities, leading to
\begin{align}
\p_t S_+ + \kappa S_+ = - \kappa E_O \delta(t+L) \\
\p_t S_- - \kappa S_- =  \kappa E_O \delta(t-L)  \ ,
\end{align}
where $E_O = h_O/\delta $. 
The delta functions are the shockwaves exiting and entering the AdS region. The signs here, and in particular the extra minus sign from the Hilbert transform, ensure that there is a sensible solution for the Schwarzian, which is time-symmetric and goes to zero as $t\to \pm \infty$. The solution is
\be
S_+ = \Theta(-t-L)
\kappa E_0 e^{\kappa(t+L)} 
\ ,  \qquad
S_- = 
\Theta(t-L) \kappa E_0 e^{\kappa(L-t)} \ .
\ee
For $t>0$, this is essentially the same solution as the evaporating black hole in \cite{Almheiri:2019psf}, which had a shockwave produced by a joining quench rather than an operator insertion. 

\small
\bibliographystyle{ourbst}
 \bibliography{replicas_draft.bib}

\providecommand{\href}[2]{#2}\begingroup\raggedright\begin{thebibliography}{10}

\bibitem{Hawking:1976ra}
S.~W. Hawking, {{Breakdown of Predictability in Gravitational Collapse}},
  \href{http://dx.doi.org/10.1103/PhysRevD.14.2460}{Phys. Rev. {\bf D14},
  2460--2473, 1976}.

\bibitem{Page:1993wv}
D.~N. Page, {{Information in black hole radiation}},
  \href{http://dx.doi.org/10.1103/PhysRevLett.71.3743}{Phys. Rev. Lett. {\bf
  71}, 3743--3746, 1993},
  [\href{http://arxiv.org/abs/arXiv:hep-th/9306083}{{arXiv:hep-th/9306083
  [hep-th]}}].

\bibitem{Page:2013dx}
D.~N. Page, {{Time Dependence of Hawking Radiation Entropy}},
  \href{http://dx.doi.org/10.1088/1475-7516/2013/09/028}{JCAP {\bf 1309}, 028,
  2013}, [\href{http://arxiv.org/abs/arXiv:1301.4995}{{arXiv:1301.4995
  [hep-th]}}].

\bibitem{Penington:2019npb}
G.~Penington, {{Entanglement Wedge Reconstruction and the Information
  Paradox}},  2019,
  [\href{http://arxiv.org/abs/arXiv:1905.08255}{{arXiv:1905.08255 [hep-th]}}].

\bibitem{Almheiri:2019psf}
A.~Almheiri, N.~Engelhardt, D.~Marolf and H.~Maxfield, {{The entropy of bulk
  quantum fields and the entanglement wedge of an evaporating black hole}},
  2019, [\href{http://arxiv.org/abs/arXiv:1905.08762}{{arXiv:1905.08762
  [hep-th]}}].

\bibitem{Almheiri:2019hni}
A.~Almheiri, R.~Mahajan, J.~Maldacena and Y.~Zhao, {{The Page curve of Hawking
  radiation from semiclassical geometry}},  2019,
  [\href{http://arxiv.org/abs/arXiv:1908.10996}{{arXiv:1908.10996 [hep-th]}}].

\bibitem{Mertens:2019bvy}
T.~G. Mertens, {{Towards Black Hole Evaporation in Jackiw-Teitelboim Gravity}},
  \href{http://dx.doi.org/10.1007/JHEP07(2019)097}{JHEP {\bf 07}, 097, 2019},
  [\href{http://arxiv.org/abs/arXiv:1903.10485}{{arXiv:1903.10485 [hep-th]}}].

\bibitem{Akers:2019wxj}
C.~Akers, A.~Levine and S.~Leichenauer, {{Large Breakdowns of Entanglement
  Wedge Reconstruction}},  2019,
  [\href{http://arxiv.org/abs/arXiv:1908.03975}{{arXiv:1908.03975 [hep-th]}}].

\bibitem{Moitra:2019xoj}
U.~Moitra, S.~K. Sake, S.~P. Trivedi and V.~Vishal, {{Jackiw-Teitelboim Model
  Coupled to Conformal Matter in the Semi-Classical Limit}},  2019,
  [\href{http://arxiv.org/abs/arXiv:1908.08523}{{arXiv:1908.08523 [hep-th]}}].

\bibitem{Almheiri:2019psy}
A.~Almheiri, R.~Mahajan and J.~E. Santos, {{Entanglement islands in higher
  dimensions}},  2019,
  [\href{http://arxiv.org/abs/arXiv:1911.09666}{{arXiv:1911.09666 [hep-th]}}].

\bibitem{Fu:2019oyc}
Z.~Fu and D.~Marolf, {{Bag-of-gold spacetimes, Euclidean wormholes, and
  inflation from domain walls in AdS/CFT}},
  \href{http://dx.doi.org/10.1007/JHEP11(2019)040}{JHEP {\bf 11}, 040, 2019},
  [\href{http://arxiv.org/abs/arXiv:1909.02505}{{arXiv:1909.02505 [hep-th]}}].

\bibitem{Zhang:2019fcy}
P.~Zhang, {{Evaporation dynamics of the Sachdev-Ye-Kitaev model}},  2019,
  [\href{http://arxiv.org/abs/arXiv:1909.10637}{{arXiv:1909.10637
  [cond-mat.str-el]}}].

\bibitem{Akers:2019nfi}
C.~Akers, N.~Engelhardt and D.~Harlow, {{Simple holographic models of black
  hole evaporation}},  2019,
  [\href{http://arxiv.org/abs/arXiv:1910.00972}{{arXiv:1910.00972 [hep-th]}}].

\bibitem{Almheiri:2019yqk}
A.~Almheiri, R.~Mahajan and J.~Maldacena, {{Islands outside the horizon}},
  2019, [\href{http://arxiv.org/abs/arXiv:1910.11077}{{arXiv:1910.11077
  [hep-th]}}].

\bibitem{Rozali:2019day}
M.~Rozali, J.~Sully, M.~Van~Raamsdonk, C.~Waddell and D.~Wakeham, {{Information
  radiation in BCFT models of black holes}},  2019,
  [\href{http://arxiv.org/abs/arXiv:1910.12836}{{arXiv:1910.12836 [hep-th]}}].

\bibitem{Chen:2019uhq}
H.~Z. Chen, Z.~Fisher, J.~Hernandez, R.~C. Myers and S.-M. Ruan, {{Information
  Flow in Black Hole Evaporation}},  2019,
  [\href{http://arxiv.org/abs/arXiv:1911.03402}{{arXiv:1911.03402 [hep-th]}}].

\bibitem{Bousso:2019ykv}
R.~Bousso and M.~Tomasevic, {{Unitarity From a Smooth Horizon?}},  2019,
  [\href{http://arxiv.org/abs/arXiv:1911.06305}{{arXiv:1911.06305 [hep-th]}}].

\bibitem{Jafferis:2019wkd}
D.~L. Jafferis and D.~K. Kolchmeyer, {{Entanglement Entropy in
  Jackiw-Teitelboim Gravity}},  2019,
  [\href{http://arxiv.org/abs/arXiv:1911.10663}{{arXiv:1911.10663 [hep-th]}}].

\bibitem{Blommaert:2019wfy}
A.~Blommaert, T.~G. Mertens and H.~Verschelde, {{Eigenbranes in
  Jackiw-Teitelboim gravity}},  2019,
  [\href{http://arxiv.org/abs/arXiv:1911.11603}{{arXiv:1911.11603 [hep-th]}}].

\bibitem{Ryu:2006bv}
S.~Ryu and T.~Takayanagi, {{Holographic derivation of entanglement entropy from
  AdS/CFT}}, \href{http://dx.doi.org/10.1103/PhysRevLett.96.181602}{Phys. Rev.
  Lett. {\bf 96}, 181602, 2006},
  [\href{http://arxiv.org/abs/arXiv:hep-th/0603001}{{arXiv:hep-th/0603001
  [hep-th]}}].

\bibitem{Hubeny:2007xt}
V.~E. Hubeny, M.~Rangamani and T.~Takayanagi, {{A Covariant holographic
  entanglement entropy proposal}},
  \href{http://dx.doi.org/10.1088/1126-6708/2007/07/062}{JHEP {\bf 07}, 062,
  2007}, [\href{http://arxiv.org/abs/arXiv:0705.0016}{{arXiv:0705.0016
  [hep-th]}}].

\bibitem{Faulkner:2013ana}
T.~Faulkner, A.~Lewkowycz and J.~Maldacena, {{Quantum corrections to
  holographic entanglement entropy}},
  \href{http://dx.doi.org/10.1007/JHEP11(2013)074}{JHEP {\bf 11}, 074, 2013},
  [\href{http://arxiv.org/abs/arXiv:1307.2892}{{arXiv:1307.2892 [hep-th]}}].

\bibitem{Engelhardt:2014gca}
N.~Engelhardt and A.~C. Wall, {{Quantum Extremal Surfaces: Holographic
  Entanglement Entropy beyond the Classical Regime}},
  \href{http://dx.doi.org/10.1007/JHEP01(2015)073}{JHEP {\bf 01}, 073, 2015},
  [\href{http://arxiv.org/abs/arXiv:1408.3203}{{arXiv:1408.3203 [hep-th]}}].

\bibitem{Mathur:2014dia}
S.~D. Mathur, {{What is the dual of two entangled CFTs?}},  2014,
  [\href{http://arxiv.org/abs/arXiv:1402.6378}{{arXiv:1402.6378 [hep-th]}}].

\bibitem{Jackiw:1984je}
R.~Jackiw, {{Lower Dimensional Gravity}},
  \href{http://dx.doi.org/10.1016/0550-3213(85)90448-1}{Nucl. Phys. {\bf B252},
  343--356, 1985}.

\bibitem{Teitelboim:1983ux}
C.~Teitelboim, {{Gravitation and Hamiltonian Structure in Two Space-Time
  Dimensions}}, \href{http://dx.doi.org/10.1016/0370-2693(83)90012-6}{Phys.
  Lett. {\bf B126}, 41--45, 1983}.

\bibitem{Almheiri:2014cka}
A.~Almheiri and J.~Polchinski, {{Models of AdS$_{2}$ backreaction and
  holography}}, \href{http://dx.doi.org/10.1007/JHEP11(2015)014}{JHEP {\bf 11},
  014, 2015}, [\href{http://arxiv.org/abs/arXiv:1402.6334}{{arXiv:1402.6334
  [hep-th]}}].

\bibitem{Lewkowycz:2013nqa}
A.~Lewkowycz and J.~Maldacena, {{Generalized gravitational entropy}},
  \href{http://dx.doi.org/10.1007/JHEP08(2013)090}{JHEP {\bf 08}, 090, 2013},
  [\href{http://arxiv.org/abs/arXiv:1304.4926}{{arXiv:1304.4926 [hep-th]}}].

\bibitem{Dong:2016hjy}
X.~Dong, A.~Lewkowycz and M.~Rangamani, {{Deriving covariant holographic
  entanglement}}, \href{http://dx.doi.org/10.1007/JHEP11(2016)028}{JHEP {\bf
  11}, 028, 2016},
  [\href{http://arxiv.org/abs/arXiv:1607.07506}{{arXiv:1607.07506 [hep-th]}}].

\bibitem{Dong:2017xht}
X.~Dong and A.~Lewkowycz, {{Entropy, Extremality, Euclidean Variations, and the
  Equations of Motion}}, \href{http://dx.doi.org/10.1007/JHEP01(2018)081}{JHEP
  {\bf 01}, 081, 2018},
  [\href{http://arxiv.org/abs/arXiv:1705.08453}{{arXiv:1705.08453 [hep-th]}}].

\bibitem{Penington:2019kki}
G.~Penington, S.~H. Shenker, D.~Stanford and Z.~Yang, {{Replica wormholes and
  the black hole interior}},  2019,
  [\href{http://arxiv.org/abs/arXiv:1911.11977}{{arXiv:1911.11977 [hep-th]}}].

\bibitem{Maldacena:2001kr}
J.~M. Maldacena, {{Eternal black holes in anti-de Sitter}},
  \href{http://dx.doi.org/10.1088/1126-6708/2003/04/021}{JHEP {\bf 04}, 021,
  2003},
  [\href{http://arxiv.org/abs/arXiv:hep-th/0106112}{{arXiv:hep-th/0106112
  [hep-th]}}].

\bibitem{Saad:2018bqo}
P.~Saad, S.~H. Shenker and D.~Stanford, {{A semiclassical ramp in SYK and in
  gravity}},  2018,
  [\href{http://arxiv.org/abs/arXiv:1806.06840}{{arXiv:1806.06840 [hep-th]}}].

\bibitem{Saad:2019lba}
P.~Saad, S.~H. Shenker and D.~Stanford, {{JT gravity as a matrix integral}},
  2019, [\href{http://arxiv.org/abs/arXiv:1903.11115}{{arXiv:1903.11115
  [hep-th]}}].

\bibitem{Saad:2019pqd}
P.~Saad, {{Late Time Correlation Functions, Baby Universes, and ETH in JT
  Gravity}},  2019,
  [\href{http://arxiv.org/abs/arXiv:1910.10311}{{arXiv:1910.10311 [hep-th]}}].

\bibitem{Rocha:2008fe}
J.~V. Rocha, {{Evaporation of large black holes in AdS: Coupling to the
  evaporon}}, \href{http://dx.doi.org/10.1088/1126-6708/2008/08/075}{JHEP {\bf
  08}, 075, 2008},
  [\href{http://arxiv.org/abs/arXiv:0804.0055}{{arXiv:0804.0055 [hep-th]}}].

\bibitem{Engelsoy:2016xyb}
J.~Engelsoy, T.~G. Mertens and H.~Verlinde, {{An investigation of AdS$_{2}$
  backreaction and holography}},
  \href{http://dx.doi.org/10.1007/JHEP07(2016)139}{JHEP {\bf 07}, 139, 2016},
  [\href{http://arxiv.org/abs/arXiv:1606.03438}{{arXiv:1606.03438 [hep-th]}}].

\bibitem{Randall:1999vf}
L.~Randall and R.~Sundrum, {{An Alternative to compactification}},
  \href{http://dx.doi.org/10.1103/PhysRevLett.83.4690}{Phys. Rev. Lett. {\bf
  83}, 4690--4693, 1999},
  [\href{http://arxiv.org/abs/arXiv:hep-th/9906064}{{arXiv:hep-th/9906064
  [hep-th]}}].

\bibitem{Karch:2000ct}
A.~Karch and L.~Randall, {{Locally localized gravity}},
  \href{http://dx.doi.org/10.1088/1126-6708/2001/05/008}{JHEP {\bf 05}, 008,
  2001},
  [\href{http://arxiv.org/abs/arXiv:hep-th/0011156}{{arXiv:hep-th/0011156
  [hep-th]}}].

\bibitem{Maldacena:2018gjk}
J.~Maldacena, A.~Milekhin and F.~Popov, {{Traversable wormholes in four
  dimensions}},  2018,
  [\href{http://arxiv.org/abs/arXiv:1807.04726}{{arXiv:1807.04726 [hep-th]}}].

\bibitem{Fiola:1994ir}
T.~M. Fiola, J.~Preskill, A.~Strominger and S.~P. Trivedi, {{Black hole
  thermodynamics and information loss in two-dimensions}},
  \href{http://dx.doi.org/10.1103/PhysRevD.50.3987}{Phys. Rev. {\bf D50},
  3987--4014, 1994},
  [\href{http://arxiv.org/abs/arXiv:hep-th/9403137}{{arXiv:hep-th/9403137
  [hep-th]}}].

\bibitem{Callan:1994py}
C.~G. Callan, Jr. and F.~Wilczek, {{On geometric entropy}},
  \href{http://dx.doi.org/10.1016/0370-2693(94)91007-3}{Phys. Lett. {\bf B333},
  55--61, 1994},
  [\href{http://arxiv.org/abs/arXiv:hep-th/9401072}{{arXiv:hep-th/9401072
  [hep-th]}}].

\bibitem{Casini:2009sr}
H.~Casini and M.~Huerta, {{Entanglement entropy in free quantum field theory}},
  \href{http://dx.doi.org/10.1088/1751-8113/42/50/504007}{J. Phys. {\bf A42},
  504007, 2009}, [\href{http://arxiv.org/abs/arXiv:0905.2562}{{arXiv:0905.2562
  [hep-th]}}].

\bibitem{Dong:2013qoa}
X.~Dong, {{Holographic Entanglement Entropy for General Higher Derivative
  Gravity}}, \href{http://dx.doi.org/10.1007/JHEP01(2014)044}{JHEP {\bf 01},
  044, 2014}, [\href{http://arxiv.org/abs/arXiv:1310.5713}{{arXiv:1310.5713
  [hep-th]}}].

\bibitem{Mumford}
E.~Sharon and D.~Mumford, {{2D-Shape Analysis Using Conformal Mapping}},
  \href{http://dx.doi.org/10.1007/s11263-006-6121-z}{International Journal of
  Computer Vision {\bf 70}, 55, 2006}.

\bibitem{Jensen:2016pah}
K.~Jensen, {{Chaos and hydrodynamics near AdS$_2$}},  2016,
  [\href{http://arxiv.org/abs/arXiv:1605.06098}{{arXiv:1605.06098 [hep-th]}}].

\bibitem{Maldacena:2016upp}
J.~Maldacena, D.~Stanford and Z.~Yang, {{Conformal symmetry and its breaking in
  two dimensional Nearly Anti-de-Sitter space}},
  \href{http://dx.doi.org/10.1093/ptep/ptw124}{PTEP {\bf 2016}, 12C104, 2016},
  [\href{http://arxiv.org/abs/arXiv:1606.01857}{{arXiv:1606.01857 [hep-th]}}].

\bibitem{Calabrese:2004eu}
P.~Calabrese and J.~L. Cardy, {{Entanglement entropy and quantum field
  theory}}, \href{http://dx.doi.org/10.1088/1742-5468/2004/06/P06002}{J. Stat.
  Mech. {\bf 0406}, P06002, 2004},
  [\href{http://arxiv.org/abs/arXiv:hep-th/0405152}{{arXiv:hep-th/0405152
  [hep-th]}}].

\bibitem{Casini:2005rm}
H.~Casini, C.~D. Fosco and M.~Huerta, {{Entanglement and alpha entropies for a
  massive Dirac field in two dimensions}},
  \href{http://dx.doi.org/10.1088/1742-5468/2005/07/P07007}{J. Stat. Mech. {\bf
  0507}, P07007, 2005},
  [\href{http://arxiv.org/abs/arXiv:cond-mat/0505563}{{arXiv:cond-mat/0505563
  [cond-mat]}}].

\bibitem{Hayden:2018khn}
P.~Hayden and G.~Penington, {{Learning the Alpha-bits of Black Holes}},  2018,
  [\href{http://arxiv.org/abs/arXiv:1807.06041}{{arXiv:1807.06041 [hep-th]}}].

\bibitem{hayden2017approximate}
P.~Hayden and G.~Penington, {Approximate quantum error correction revisited:
  Introducing the alpha-bit},  2017.

\bibitem{Jafferis:2015del}
D.~L. Jafferis, A.~Lewkowycz, J.~Maldacena and S.~J. Suh, {{Relative entropy
  equals bulk relative entropy}},
  \href{http://dx.doi.org/10.1007/JHEP06(2016)004}{JHEP {\bf 06}, 004, 2016},
  [\href{http://arxiv.org/abs/arXiv:1512.06431}{{arXiv:1512.06431 [hep-th]}}].

\bibitem{Dong:2016eik}
X.~Dong, D.~Harlow and A.~C. Wall, {{Reconstruction of Bulk Operators within
  the Entanglement Wedge in Gauge-Gravity Duality}},
  \href{http://dx.doi.org/10.1103/PhysRevLett.117.021601}{Phys. Rev. Lett. {\bf
  117}, 021601, 2016},
  [\href{http://arxiv.org/abs/arXiv:1601.05416}{{arXiv:1601.05416 [hep-th]}}].

\bibitem{Almheiri:2014lwa}
A.~Almheiri, X.~Dong and D.~Harlow, {{Bulk Locality and Quantum Error
  Correction in AdS/CFT}},
  \href{http://dx.doi.org/10.1007/JHEP04(2015)163}{JHEP {\bf 04}, 163, 2015},
  [\href{http://arxiv.org/abs/arXiv:1411.7041}{{arXiv:1411.7041 [hep-th]}}].

\bibitem{Strominger:1996sh}
A.~Strominger and C.~Vafa, {{Microscopic origin of the Bekenstein-Hawking
  entropy}}, \href{http://dx.doi.org/10.1016/0370-2693(96)00345-0}{Phys. Lett.
  {\bf B379}, 99--104, 1996},
  [\href{http://arxiv.org/abs/arXiv:hep-th/9601029}{{arXiv:hep-th/9601029
  [hep-th]}}].

\bibitem{Hawking:1987mz}
S.~W. Hawking, {{Quantum Coherence Down the Wormhole}},
  \href{http://dx.doi.org/10.1016/0370-2693(87)90028-1}{Phys. Lett. {\bf B195},
  337, 1987}.

\bibitem{Lavrelashvili:1987jg}
G.~V. Lavrelashvili, V.~A. Rubakov and P.~G. Tinyakov, {{Disruption of Quantum
  Coherence upon a Change in Spatial Topology in Quantum Gravity}}, {JETP Lett.
  {\bf 46}, 167--169, 1987}.

\bibitem{Giddings:1987cg}
S.~B. Giddings and A.~Strominger, {{Axion Induced Topology Change in Quantum
  Gravity and String Theory}},
  \href{http://dx.doi.org/10.1016/0550-3213(88)90446-4}{Nucl. Phys. {\bf B306},
  890--907, 1988}.

\bibitem{Coleman:1988cy}
S.~R. Coleman, {{Black Holes as Red Herrings: Topological Fluctuations and the
  Loss of Quantum Coherence}},
  \href{http://dx.doi.org/10.1016/0550-3213(88)90110-1}{Nucl. Phys. {\bf B307},
  867--882, 1988}.

\bibitem{Giddings:1988cx}
S.~B. Giddings and A.~Strominger, {{Loss of Incoherence and Determination of
  Coupling Constants in Quantum Gravity}},
  \href{http://dx.doi.org/10.1016/0550-3213(88)90109-5}{Nucl. Phys. {\bf B307},
  854--866, 1988}.

\bibitem{Polchinski:1994zs}
J.~Polchinski and A.~Strominger, {{A Possible resolution of the black hole
  information puzzle}}, \href{http://dx.doi.org/10.1103/PhysRevD.50.7403}{Phys.
  Rev. {\bf D50}, 7403--7409, 1994},
  [\href{http://arxiv.org/abs/arXiv:hep-th/9407008}{{arXiv:hep-th/9407008
  [hep-th]}}].

\bibitem{Hartman:2013qma}
T.~Hartman and J.~Maldacena, {{Time Evolution of Entanglement Entropy from
  Black Hole Interiors}}, \href{http://dx.doi.org/10.1007/JHEP05(2013)014}{JHEP
  {\bf 05}, 014, 2013},
  [\href{http://arxiv.org/abs/arXiv:1303.1080}{{arXiv:1303.1080 [hep-th]}}].

\bibitem{Susskind:2014moa}
L.~Susskind, {{Entanglement is not enough}},
  \href{http://dx.doi.org/10.1002/prop.201500095}{Fortsch. Phys. {\bf 64},
  49--71, 2016}, [\href{http://arxiv.org/abs/arXiv:1411.0690}{{arXiv:1411.0690
  [hep-th]}}].

\end{thebibliography}\endgroup
\end{document}